\documentclass[prd,aps,floats,twocolumn,nofootinbib]{revtex4-1}

\usepackage[utf8]{inputenc}
\usepackage[T1]{fontenc}
\usepackage{amsmath}
\usepackage{amsfonts}
\usepackage{amssymb}
\usepackage{graphicx}
\usepackage{lmodern}
\usepackage{physics}
\usepackage[left=2cm,right=2cm,top=2cm,bottom=2cm]{geometry}
\usepackage{siunitx}
\usepackage{fancyhdr}
\usepackage{enumerate}
\usepackage{mhchem}
\usepackage{mathtools}
\usepackage{graphicx}
\usepackage{float}
\usepackage{bm}
\newcommand{\plk}{\mathfrak{h}}
\newcommand{\be}{\begin{equation}}
\newcommand{\ee}{\end{equation}}
\newcommand{\bea}{\begin{eqnarray}}
\newcommand{\eea}{\end{eqnarray}}

\newcommand{\oarX}[1]{\href{http://arxiv.org/abs/#1}{{\ttfamily #1}}}
\newcommand{\arX}[1]{\href{http://arxiv.org/abs/#1}{{\ttfamily arXiv:#1}}}

\DeclareSIUnit\year{a}
\DeclareSIUnit\cliVht{c}

\begin{document}

\title{Possible quantum effects at the transition from cosmological deceleration to acceleration
}
\author{Bruno Alexandre}
\author{Jo\~{a}o Magueijo}
\email{magueijo@ic.ac.uk}
\affiliation{Theoretical Physics Group, The Blackett Laboratory, Imperial College, Prince Consort Rd., London, SW7 2BZ, United Kingdom}

\date{\today}

\begin{abstract}
The recent transition from decelerated to accelerated expansion can be seen as a reflection (or ``bounce'') in the connection variable,  
defined by the inverse comoving Hubble length ($b=\dot a$, on-shell). We study the quantum cosmology 
of this process. We use a formalism for obtaining relational time variables either through the demotion of the constants of Nature to integration constants, or by identifying fluid constants of motion. We  extend its previous application to a toy model (radiation and $\Lambda$) to the realistic setting of a transition from dust matter to $\Lambda$ domination.
In the dust and $\Lambda$ model two time variables may be defined, conjugate to $\Lambda$  and  to the dust constant of motion, and we work out the monochromatic solutions to the Schr\"odinger equation representing the Hamiltonian constraint. As for their radiation and $\Lambda$ counterparts, these solutions exhibit ``ringing'', whereby the incident and reflected waves interfere,
leading to oscillations in the amplitude. In the semi-classical approximation we find that, close to the bounce, the probability distribution becomes double-peaked, one peak following a trajectory close to the classical limit but with a Hubble parameter slightly shifted  downwards, the other with a value of $b$ stuck at its minimum $b=b_\star$.
Still closer to the transition, the distribution is better approximated by an exponential distribution, with a single peak at $b=b_\star$, and a (more representative) average $b$  biased towards a value higher than the classical trajectory. Thus, we obtain a distinctive prediction for the average Hubble parameter with redshift: slightly lower than its classical value when $z\approx 0$, but potentially much higher than the classical prediction around $z\sim 0.64$, where the bounce most likely occurred. The implications for the ``Hubble tension'' have not escaped us.  
\end{abstract}

\maketitle

\section{Introduction}

The connection representation is sometimes favoured over the metric representation in quantum gravity. This is the case in loop quantum gravity, but the issue is more general. The choice of representation is non-trivial and can lead to non-equivalent quantum theories due to matters of ordering, inner product, boundary conditions, etc. The choice of representation also unveils new interpretations of some phenomena. For example, the well-established (\cite{accelexp} and references therein) recent transition from decelerated to accelerated expansion can be seen as a reflection or bounce in the cosmological connection variable (this is not to be confused with a primordial  bounce in the metric $a$). The connection variable associated with homogenous cosmological models is the inverse comoving Hubble parameter (herein $b$), with $b=\dot a/N$ on-shell (where $a$ is the expansion factor and $N$ is the lapse function). This is precisely
the variable used in characterizing the horizon structure of the Universe: it decreases in time for decelerated expansion, increases in time for accelerated expansion. Hence, $b$ must have reflected or bounced off a minimum $b=b_*$ in our recent past (around redshift $z\sim 0.64$ in most models).

Reflection can bring to the fore quantum  behavior~\cite{interfreflex}, sometimes paradoxically~\cite{quantumreflex}. The incident and reflected waves interfere, inducing oscillations in the probability (``ringing'''), spoiling the semiclassical limit. The turning point, dividing classically allowed and forbidden regions, is always surrounded by a region where the  semiclassical limit 
breaks down, revealing fully quantum behavior. This matter was given a preliminary investigation in~\cite{gielen} in the context of a toy model exhibiting reflection in $b$: a Universe containing a mixture of radiation and $\Lambda$. Here we extend this work to the realistic but technically more demanding case of a transition from matter (dust) domination to $\Lambda$ domination, as witnessed by our Universe 
some 7 billion years.

As in~\cite{gielen} we use a generalization of unimodular gravity~\cite{unimod1} as formulated in~\cite{unimod}. In such theories a relational time (converting the Wheeler-DeWitt equation into a 
Schr\"odinger-like equation) is obtained by demoting a constant of Nature to a constant-on-shell only~\cite{JoaoLetter,JoaoPaper} (i.e., a
quantity which is constant as a result of the equations of motion, rather than being a fixed parameter in the action). The conjugates
of such ``constants'' supply excellent physical time variables. In the case of $\Lambda$ this is the 4-D volume to the past of the observer~\cite{unimod}. Extensions targeting other constants (for example Newton's constant) have been considered before, for example in the sequester~\cite{padilla, pad,pad1} (where the associated ``times'' are called ``fluxes''), or  in~\cite{vikman,vikman1}. The approach of~\cite{GielenMenendez} produces equivalent results using instead fluid constants of motion and their conjugates, as defined in the Lagrangian formulation of~\cite{Brown,GielenTurok,GielenMenendez}. In a sense the constants of motion and the deconstantized constants are put on the same footing in this approach.

As shown in~\cite{GielenMenendez,JoaoPaper,twotimes,sing}, in this approach to the Wheeler--DeWitt (WDW) equation, one finds that the fixed constant solutions appear as mono-chromatic partial waves. The general
solution is a superposition of such partial waves, with amplitudes that depend on the no-longer-constants. Such superpositions can form wave packets with better normalizability properties. The first purpose of this paper is to work out these solutions, proceeding as follows. 

In Section~\ref{classical} we first formulate the classical theory in the same format as that discussed in~\cite{gielen}, but adapted to a mixture of $\Lambda$ and dust matter. Whereas before we had a quadratic equation with two solutions, we now find a cubic with a spurious solution (with $a<0$) in addition to two solutions corresponding to matter and $\Lambda$ domination. What follows is a carbon copy of~\cite{gielen}, albeit with significant further technical hurdles.  We find the solutions to the WDW equation in Section~\ref{monochrome}. In spite of their much more complicated expressions, they are, just as in~\cite{gielen}, composed of an incident and a reflected wave for $b>b_\star$, and an evanescent wave for $b<b_\star$, which we combine using matching conditions at $b=b_\star$. In Section~\ref{wavepackets} we then augment these solutions with a time-factor, to produce the monochromatic partial waves, and superpose these in wave packets. 

The second purpose of this paper is to apply these results to the real world, which we do in Section~\ref{phenomenology}.
We focus on corrections within the semiclassical limit for the probability measure, and use the average $b$ as the relevant quantity
for making predictions for departures from the classical trajectory. The results depend on the variance chosen for the coherent state
representing the wave function of the Universe. It is possible that the Lambda clock only becomes a suitable clock around the start of the Lambda 
epoch. The fact that we have a reflection makes the average $b$ be higher than the classical trajectory, and we evaluate the profile of this
effect with redshift.

\section{The classical theory}\label{classical}

We consider a model with two fluids: $\Lambda$ or dark energy ($w=-1$), and pressure-less or dust matter ($w=0$). Hence, the Hamiltonian constraint is classically equivalent to:
\begin{eqnarray}
-b^2-k+\frac{\Lambda}{3}a^2+\frac{m}{a}=0
\end{eqnarray}
which can be written in the form of a cubic equation:
\begin{eqnarray}
a^3-V(b)\phi a+m\phi=0,
\label{eqa}
\end{eqnarray}
where $V(b)=b^2+k$ and $\phi=\frac{3}{\Lambda}$. In analogy with the quadratic found in~\cite{gielen} this is the equation we have to solve in order to implement a multi-branch connection representation of the WDW equation. Being a cubic equation, we know that it has 3 solutions, which can either be all real, or one be real and the other two complex conjugates. These two cases depend on the sign of discriminant of the equation:
\be\label{Delta}
\Delta=V^3-\frac{27 m^2}{4\phi}.
\ee
It is helpful to understand the physical nature of the problem in order to select the correct solutions.

\subsection{Some physical guidance}
As with the case of $\Lambda$ and radiation studied in~\cite{gielen}, an expanding Universe will start by decelerating (in this case due to matter domination), to then transition to accelerated expansion and $\Lambda$ domination. Hence,
there will be a ``bounce'' in $b$ (the inverse comoving Hubble length):
$b$ decreases at first, when matter dominates, to bounce off a minimum value $b_*$ as $\Lambda$ creeps in, and then continue increasing as $\Lambda$ dominates. The classically allowed region is $b\ge b_*$ and this should have {\it two} branches, one containing the matter dominated solution, the other the $\Lambda$ dominated solution. The $0<b<b_*$ region is classically forbidden (we will only be considering expanding Universes in this paper, so $b>0$).

Considering the Friedmann equations:
\bea
b^2+k=V&=&\frac{a^2}{\phi}+\frac{m}{a}\label{F1}\\
\dot b&=&\frac{a}{\phi}-\frac{m}{2a^2}\label{F2}
\eea
we see that (\ref{F2}) implies that the bounce in $b$ (i.e. $\dot b=0$) happens when: 
\be
a=a_*=\left(\frac{m\phi}{2}\right)^{1/3} \label{eqastar}
\ee
with (\ref{F1}) then implying:
\be
V^3=V_*^3=(b_*^2+k)^3=\frac{27 m^2}{4\phi} \label{eqbstar}
\ee
(which defines $b_*$; for $k=0$, we have $b_\star^6=27m^2/(4\phi)$).
Comparing with (\ref{Delta}) we note that 
the bounce point is the point where the discriminant $\Delta$ vanishes. We can therefore wed the mathematics and physics as follows. 

For $\Delta>0$ we are  in the classically allowed
region, and we know that when 
\be
V^3\gg \frac{27 m^2}{4\phi}=V_*^3
\ee
(i.e. asymptotically)
there are two branches, one corresponding to matter domination, the other to $\Lambda$ domination. Hence for $\Delta>0$ we expect 3 real roots, one of which negative (i.e. $a<0$), so that it can be discarded as non-physical. The other two should be positive, one representing the branch with $\Lambda$ domination, the other the branch with matter domination. They become degenerate (a double real positive root) at $b=b_*$. 

For the classically forbidden region we expect the real root to be negative (indeed a continuation of the negative root in the allowed region, which we have discarded). The other two roots must be complex conjugates, with a real part connecting to the degenerate positive roots found at $b=b_*$. Their imaginary parts will be symmetric, so that quantum mechanically one will describe an evanescent wave, the other a wave blowing up exponentially, to be discarded.  

These physical arguments will help us understand the mathematics that follows, as well as select the correct solutions.

\subsection{Solving the algebraic equation}
\begin{figure}[H]
\centering
\includegraphics[scale=0.5]{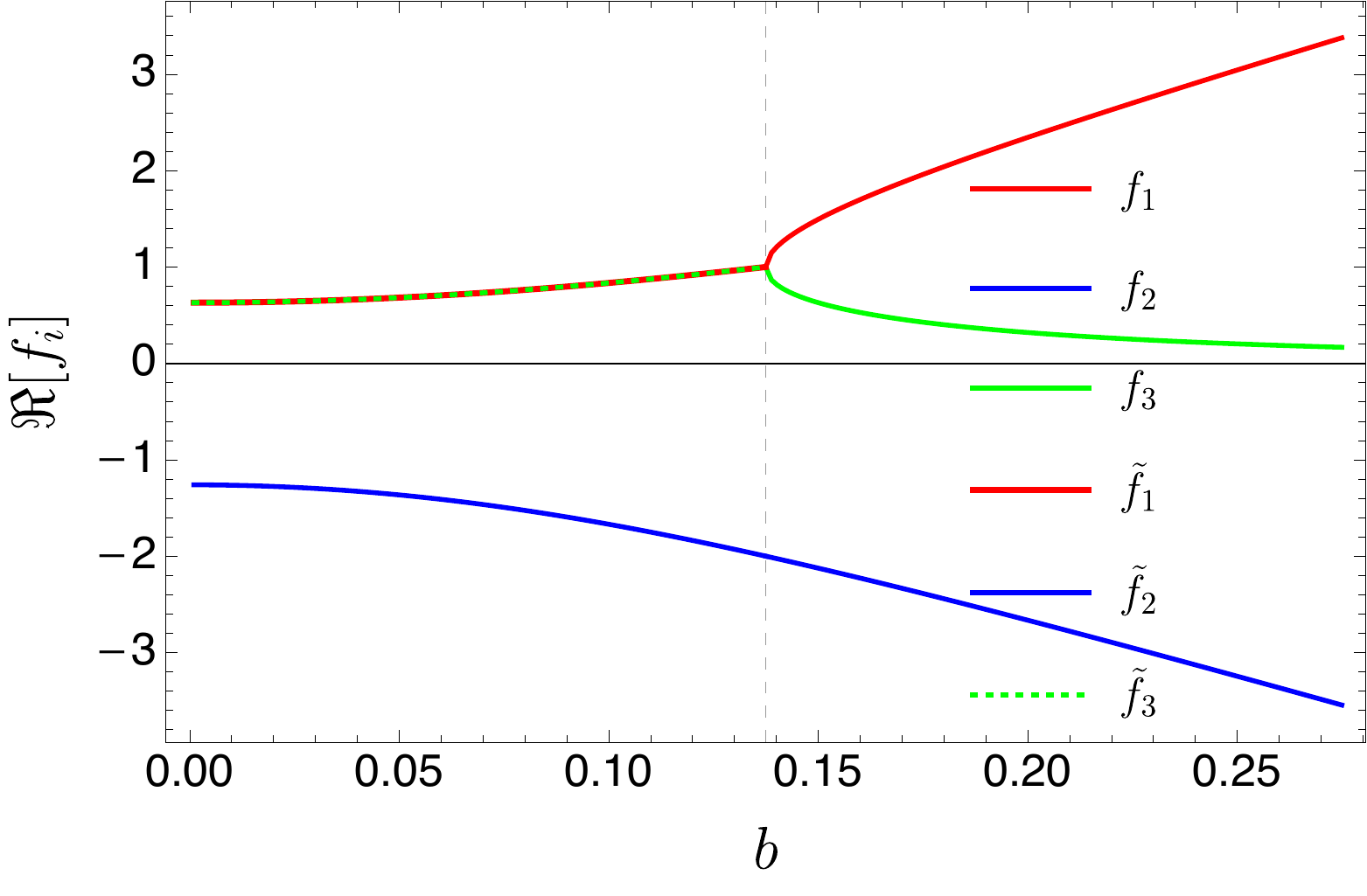}
\includegraphics[scale=0.5]{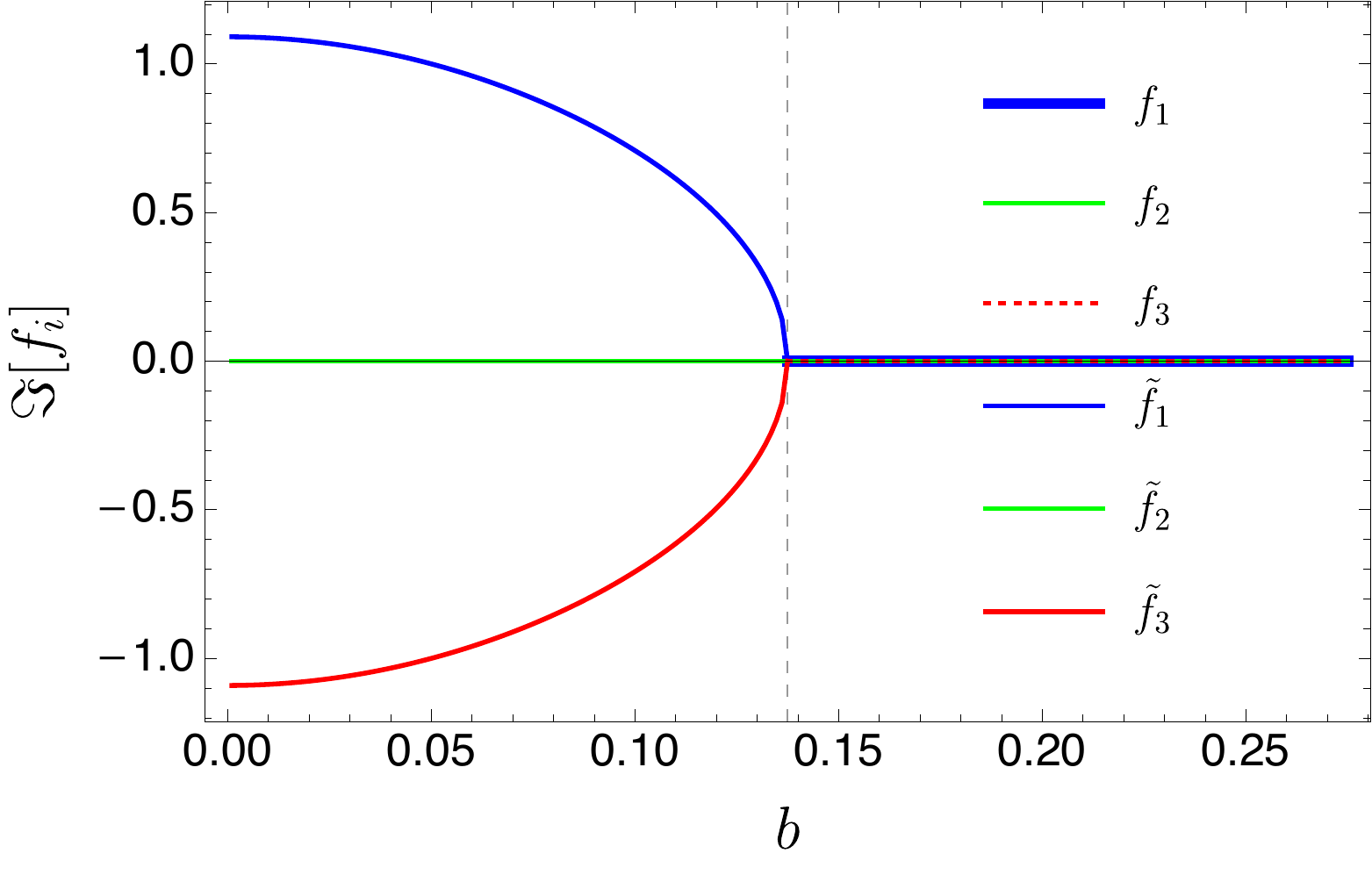}
\caption{The real and imaginary parts of the solutions of the cubic equation (\ref{eqa}) as a function of $b$ for $m=1$, $\phi=10^6$ and $k=0$. The vertical dashed line separates the classically allowed region $b>b_*$ from the forbidden region $b<b_*$. We see how $f_1$ and $f_3$ represent the physical double branches. }
\label{roots}
\end{figure}

Solving equation (\ref{eqa}) for $a$ we obtain three roots, whose form will depend on the region we are considering. 
For $V\ge V_*$ we have $3$ real solutions given by 
\begin{eqnarray}
&& a_1=a_*[(-1+ix)^{1/3}+(-1-ix)^{1/3}] \nonumber\\
&& a_2=-a_*\left[\left(-1+ix\right)^{1/3}e^{-i\pi/3}+\left(-1-ix\right)^{1/3}e^{i\pi/3}\right] \nonumber\\
&& a_3=-a_*\left[\left(-1+ix\right)^{1/3}e^{i\pi/3}+\left(-1-ix\right)^{1/3}e^{-i\pi/3}\right],\nonumber
\end{eqnarray}
with $x=\sqrt{\frac{V^3}{V_*^3}-1}$. To avoid issues with conventions for negative roots, these expressions can be written in the more explicit form:  
\begin{eqnarray}
&& a_1=2a_*(1+x^2)^{1/6}\cos(\frac{\arctan x}{3}-\frac{\pi}{3})\equiv a_*f_1 \\
&& a_2=-2a_*(1+x^2)^{1/6}\cos(\frac{\arctan x}{3})\equiv a_*f_2 \\
&& a_3=2a_*(1+x^2)^{1/6}\cos(\frac{\arctan x}{3}+\frac{\pi}{3})\equiv a_*f_3.
\end{eqnarray}
Taking the asymptotic limit $V\gg V_*$ we do recover the $\Lambda$ and matter dominated solutions for $a_1$ and $a_3$ (respectively): 
\begin{eqnarray}
&& a_1\approx\sqrt{V\phi} \\
&& a_3\approx\frac{m}{V}.
\end{eqnarray}
The second root is negative, as expected, and so can be discarded. We illustrate this behaviour in Fig.\ref{roots}, right hand side of the plots, where the classically allowed region lies. 

For $V<V_*$ we obtain 1 real and 2 complex solutions (which are the complex conjugate of each other):
\begin{eqnarray}
&& a_1=2a_*(1-\Tilde{x}^2)^{1/6}\cos(\frac{\arctan i\Tilde{x}}{3}-\frac{\pi}{3}) \nonumber\\
&& a_2=-2a_*(1-\Tilde{x}^2)^{1/6}\cos(\frac{\arctan i\Tilde{x}}{3}) \nonumber\\
&& a_3=2a_*(1-\Tilde{x}^2)^{1/6}\cos(\frac{\arctan i\Tilde{x}}{3}+\frac{\pi}{3}),\nonumber
\end{eqnarray}
with $x=i\Tilde{x}$ and $\Tilde{x}=\sqrt{1-\frac{V^3}{V_*^3}}$. Using the identity
\begin{eqnarray}
\arctan(i\Tilde{x})=\frac{i}{2}\ln\frac{1+\Tilde{x}}{1-\Tilde{x}}
\end{eqnarray}
one can write these roots as:
\begin{widetext}
\begin{eqnarray}
&& a_1=\frac{a_*}{2}(1-\Tilde{x}^2)^{1/6}\left[\left(\frac{1-\Tilde{x}}{1+\Tilde{x}}\right)^{1/6}+\left(\frac{1+\Tilde{x}}{1-\Tilde{x}}\right)^{1/6}+i\sqrt{3}\left(\left(\frac{1+\Tilde{x}}{1-\Tilde{x}}\right)^{1/6}-\left(\frac{1-\Tilde{x}}{1+\Tilde{x}}\right)^{1/6}\right)\right]\equiv a_*\Tilde{f}_1 \\
&& a_2=-a_*(1-\Tilde{x}^2)^{1/6}\left[\left(\frac{1-\Tilde{x}}{1+\Tilde{x}}\right)^{1/6}+\left(\frac{1+\Tilde{x}}{1-\Tilde{x}}\right)^{1/6}\right]\equiv a_*\Tilde{f}_2 \\
&& a_3=\frac{a_*}{2}(1-\Tilde{x}^2)^{1/6}\left[\left(\frac{1-\Tilde{x}}{1+\Tilde{x}}\right)^{1/6}+\left(\frac{1+\Tilde{x}}{1-\Tilde{x}}\right)^{1/6}-i\sqrt{3}\left(\left(\frac{1+\Tilde{x}}{1-\Tilde{x}}\right)^{1/6}-\left(\frac{1-\Tilde{x}}{1+\Tilde{x}}\right)^{1/6}\right)\right]\equiv a_*\Tilde{f}_3 ,
\end{eqnarray}
\end{widetext}
showing explicitly that the real solution is $a_2$ and continues to be negative (and so to be discarded, as before), whereas the other solutions 
are the conjugate complex solutions. These features are displayed in the left hand side of Fig.~\ref{roots}, corresponding to the classically forbidden region. Note how $a_1$ and $a_3$ converge to $a_*$ for $V=V_*^+$, the point for which we have a bounce in $b$. 

As a side remark, we note that this two-branch structure could be an expression in the connection representation language of the issues raised in~\cite{twosheet}.

\section{The monochromatic solutions}\label{monochrome}
A solution to the WDW equation in the connection representation can now be found just as in~\cite{gielen}, with the proviso that we should discard one of the solutions of the cubic equation.  Hence we start from solutions $a-a_n=0$ with $n=1,3$, for which we shall now relabel to $n=\pm$, where $-=1$ and $+=3$. We also write $a_{\pm}=a_*F_{\pm}$ with $F_{\pm}$ defined as
$f_{\pm}$ for $b>b_*$ and $\Tilde{f}_{\pm}$ for $b<b_*$ (as defined in the previous Section). The Hamiltonian constraint can then be put in the form proposed in~\cite{gielen,JoaoPaper}:
\begin{eqnarray}\label{ham}
h_{\pm}(b)a^2-m^2=0
\end{eqnarray}
where $h_{\pm}(b)$ is given by
\begin{eqnarray}
h_{\pm}(b)=\left[\left(\frac{\phi}{2m^2}\right)^{2/3}F_{\pm}^2\right]^{-1}.
\end{eqnarray}
Upon quantization we get the WDW equation in the connection representation:
\begin{eqnarray}
\left(-i\plk h_{\pm}(b)\frac{\partial}{\partial b}-m^2\right)\psi_{s\pm}=0
\label{eqwdw}
\end{eqnarray}
with $\plk=\frac{l_P^3}{3V_c}$, where $V_c$ is the comoving spatial volume of the region under study ($V_c=2\pi^2$ for a whole $k=1$ Universe), $l_P=\sqrt{8\pi G_0\hbar}$ is the reduced Planck length and $G_0$ is the gravitational constant.

\begin{figure}[H]
\centering
\includegraphics[scale=0.53]{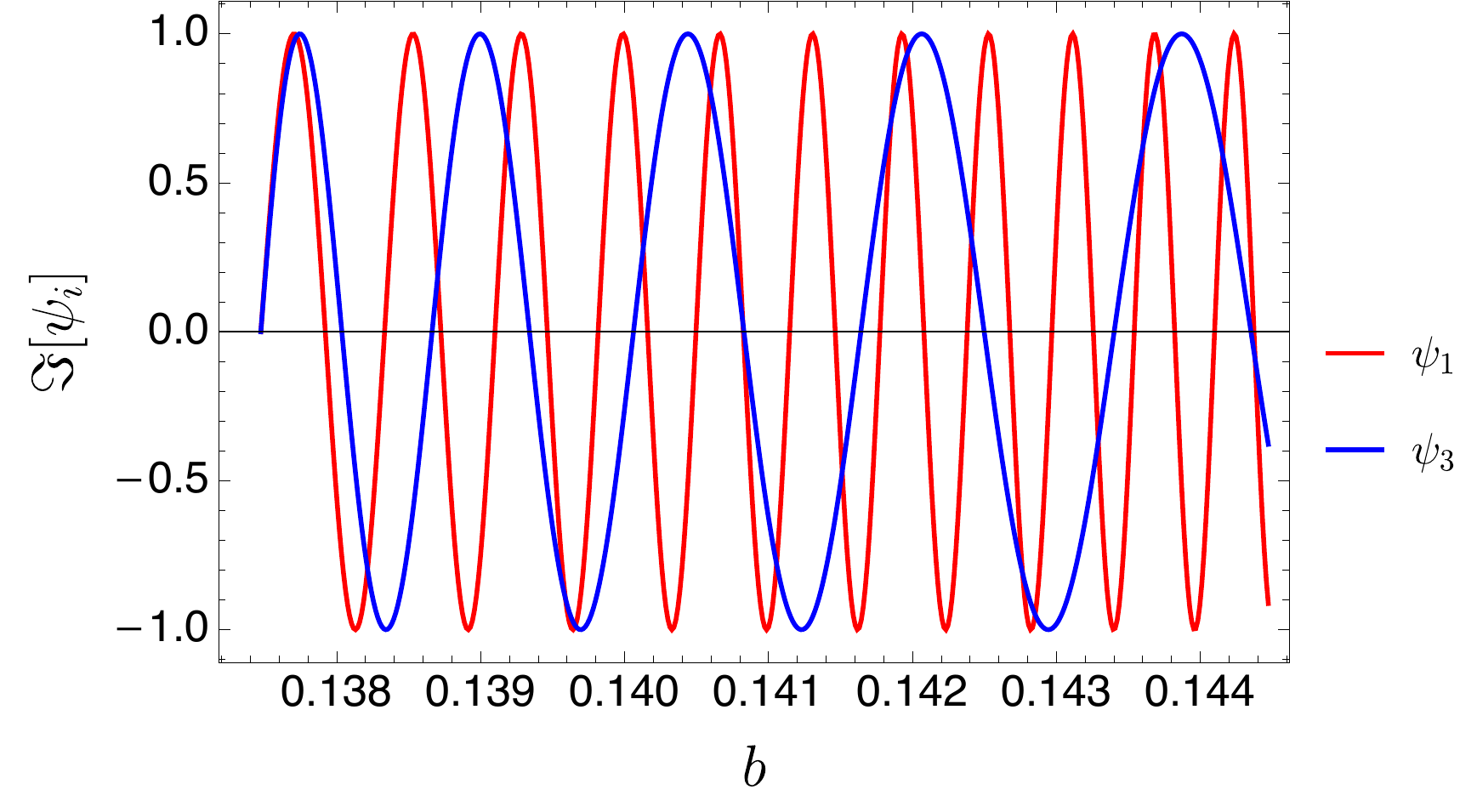}
\includegraphics[scale=0.5]{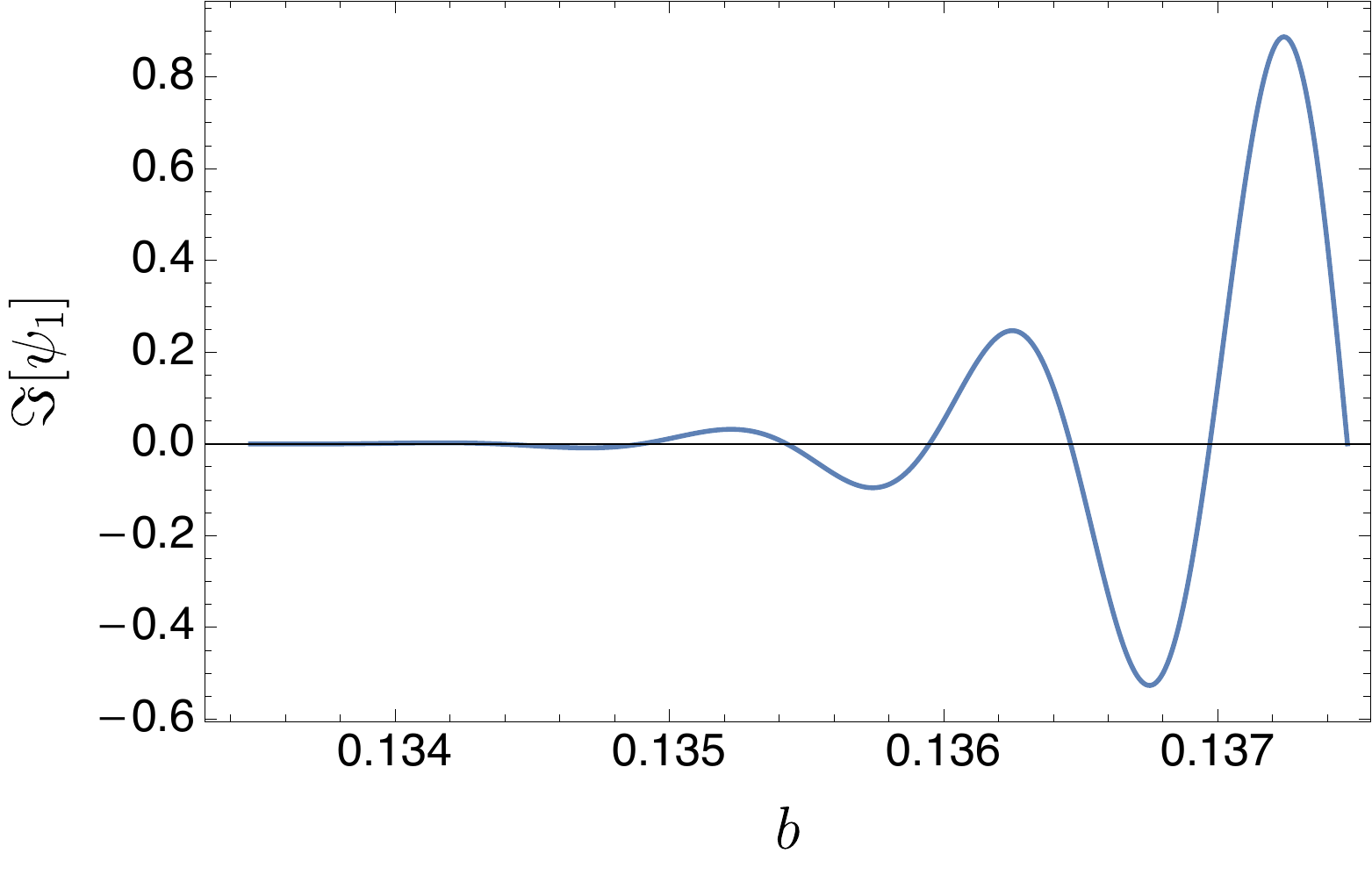}
\caption{Imaginary part of the wave functions associated with the solutions $a_1$ and $a_3$ in the classical (top panel) and forbidden (bottom panel) regions for $k=0$, $m=1$ and $\phi=10^6$, as a function of $b$.} 
\label{fig2}
\end{figure}

As in~\cite{gielen} we can define a linerizing variable:
\begin{eqnarray}
X_{\pm}(b)=\int\frac{db}{h_{\pm}(b)}=\left(\frac{\phi}{2m^2}\right)^{2/3}\int dbF_{\pm}^2.
\end{eqnarray}
which generalizes the Chern-Simons functional. For the rest of this paper we shall assume $k=0$. 
%
%
We can then simplify its expression as:
\begin{eqnarray}
X_{\pm}=\frac{4\phi}{3m^2}\int_{b_\star} db\,  b^2\cos^2\left(\frac{\arctan x}{3}\pm\frac{\pi}{3}\right).
\end{eqnarray}
(since for $k=0$, $V_\star=b_\star^2$). The solutions of (\ref{eqwdw}) are linear combinations of plane waves in the linearizing variables $X_{\pm}$, which will depend on the region of $b$ space that we are considering. For $b>b_*$ we find a superposition of an incident and a reflected wave:
\begin{eqnarray}
\psi_{s>}=A_-e^{\frac{i}{\plk}m^2X_-}+A_+e^{\frac{i}{\plk}m^2X_+}.
\end{eqnarray}
For $b<b_*$ we have an evanescent and a divergent solution. Rejecting the latter we  find:
\begin{eqnarray}
\psi_{s<}=Be^{\frac{i}{\plk}m^2\Tilde{X}_-}.
\end{eqnarray}
where $\Tilde{X}$ corresponds to the case of $\Tilde{f}$. A plot of the imaginary part of these wave functions for both regions is shown in figure \ref{fig2}.

\begin{figure}[H]
\centering
\includegraphics[scale=0.5]{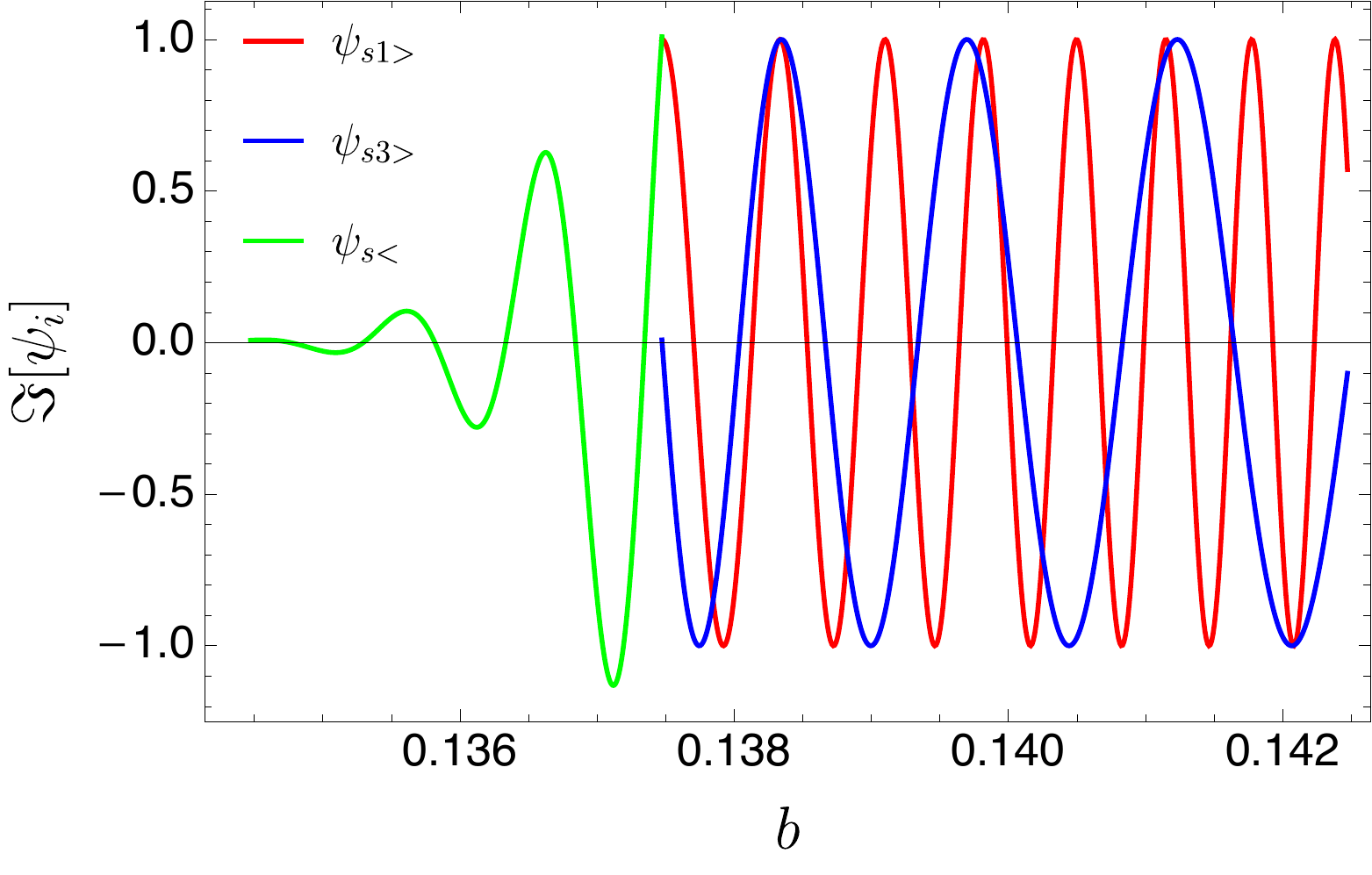}
\caption{Imaginary part of the wave functions in both regions after considering the boundary conditions in $b_*$, for $k=0$, $m=1$ and $\phi=10^6$.} 
\label{fig3}
\end{figure}

Thus, we have to solve a problem of wave reflection in quantum mechanics. One should match the wave functions in the bouncing point $b=b_*$, as well as their first and second derivatives. The first boundary condition leads to:
\begin{eqnarray}
\psi_{s>}(b_\star)=\psi_{s<}(b_\star) 
\Rightarrow
&& A_-+A_+=B.
\end{eqnarray}
For the first derivatives one gets the same condition as above and for the second we should match the divergent terms which lead to:
\begin{eqnarray}
&& B=i(A_--A_+).
\end{eqnarray}
Combining these relations we can solve the system in terms of one free parameter, which we choose to be $A_+$, and consequently fix all the other coefficients.  Therefore we arrive at
\begin{eqnarray}
A_-=-\frac{1+i}{1-i}A_+.
\end{eqnarray}
In this way the wave functions are fully determined in both regions as we have the freedom to fix $A_+$ as we see fit, which in this case is picking a phase factor that gives the correct matter limit $V>>V_*$ for $\psi_+$.
The imaginary part of the full wave function is represented in figure \ref{fig3}, where we can see both the evanescent and the propagating waves in each region. For the rest of this paper we will focus on the propagating solutions in the $b>b_\star$ region. 

\hspace{1pt}

\section{Wave packets and time evolution}\label{wavepackets}

Following~\cite{JoaoPaper} and still mimicking its implementation in~\cite{gielen}, we now construct superpositions evolving in times conjugate to a set of target constant $\boldsymbol{\alpha}$, generalizing the procedure for unimodular gravity. 
We choose:
\be
\boldsymbol{\alpha}=(m^2,\phi)
\ee
in order to recover the mono-fluid solutions in the 
 appropriate limits for matter and Lambda domination. 
Thus, we arrive at:
\begin{eqnarray}
\psi_{\pm}(b,\boldsymbol{T})=\int d\boldsymbol{\alpha}A(\boldsymbol{\alpha})e^{-\frac{i}{\plk}\boldsymbol{\alpha}\cdot\boldsymbol{T}}\psi_{s\pm}(b,\boldsymbol{\alpha}),
\end{eqnarray}
where the ``spatial'' part 
$\psi_{s\pm}$ was found in the last Section. These superpositions have simple approximate solutions if the amplitudes are sufficiently peaked. Defining: 
\begin{eqnarray}
P_{\pm}=m^2X_{\pm}(b,{\boldsymbol\alpha)}
\end{eqnarray}
so that $\psi_{s\pm}=\exp[\frac{i}{\plk}P_\pm]$,
we proceed in the same way as in \cite{gielen} using the saddle point approximation for a dispersive medium. It leads to
\begin{eqnarray}
\psi_{\pm}(b,\boldsymbol{T})\approx e^{\frac{i}{\plk}(P_{\pm}(b,\boldsymbol{\alpha}_0)-\boldsymbol{\alpha}_0\cdot\boldsymbol{T})}\prod_i\psi_{\pm i}(b,T_i)
\label{psipmap}
\end{eqnarray}
where the envelopes have expressions 
\begin{widetext}
\begin{eqnarray}
\psi_{\pm i}(b,T_i)=\int \frac{d\alpha_i}{\sqrt{2\pi\plk}}A(\alpha_i)\exp\left[-\frac{i}{\plk}(\alpha_i-\alpha_{i0})\left(T_i-X^{\text{eff}}_{\pm i}(b)\right)\right]
\label{psii}
\end{eqnarray}
\end{widetext}
with effective generalizations of the Chern-Simons functional:
\begin{eqnarray}
X^{\text{eff}}_{\pm i}(b)=\frac{\partial P_{\pm}}{\partial\alpha_i}\biggr\rvert_{\boldsymbol{\alpha}_0}.
\end{eqnarray}
Specializing to wave packets with Gaussian amplitudes $A(\alpha_i)$:
\begin{eqnarray}
A(\alpha_i)=\frac{1}{(2\pi\sigma_i^2)^{1/4}}\exp\left[-\frac{(\alpha_i-\alpha_{i0})^2}{4\sigma_i^2}\right],
\end{eqnarray}
equation (\ref{psii}) reduces to a complementary Gaussian:
\begin{eqnarray}
\psi_{\pm i}(b,T_i)=\frac{1}{(2\pi\sigma_{Ti}^2)^{1/4}}\exp\left[-\frac{(X^{\text{eff}}_{\pm i}(b)-T_i)^2}{4\sigma_{Ti}^2}\right],
\end{eqnarray}
with:
\be
\sigma_{Ti}=\frac{\plk}{2\sigma_i}.
\ee
Thus, this is an example in this context of a coherent state, i.e.
a state 
saturating the Heisenberg relations between the constants and their times.
The full time-dependent wave function in the classical region ($b>b_\star$) is of the form:
\begin{equation}
\psi_>(b,\boldsymbol{T})=A_+\psi_+(b,\boldsymbol{T})+A_-\psi_-(b,\boldsymbol{T}).
\label{eqpsib}
\end{equation}
i.e. it is made of an incident and a reflected wave, each of which comprises two factors, one for each clock being used.

It should be noted that here, just as for the radiation plus $\Lambda$ model~\cite{JoaoPaper,gielen}, the classical trajectory is given by:
\be
\dot X^{\rm eff }_{\pm i }=\dot T_i
\ee
(with $i=m,\phi$)
as we have succeeded in proving numerically (the algebra proves forbidding). Hence the peak of the Gaussian wave packets follows the classical trajectory for both factors, with either clock.

\subsection{MATTER CLOCK}
\begin{figure}[H]
\centering
\includegraphics[scale=0.53]{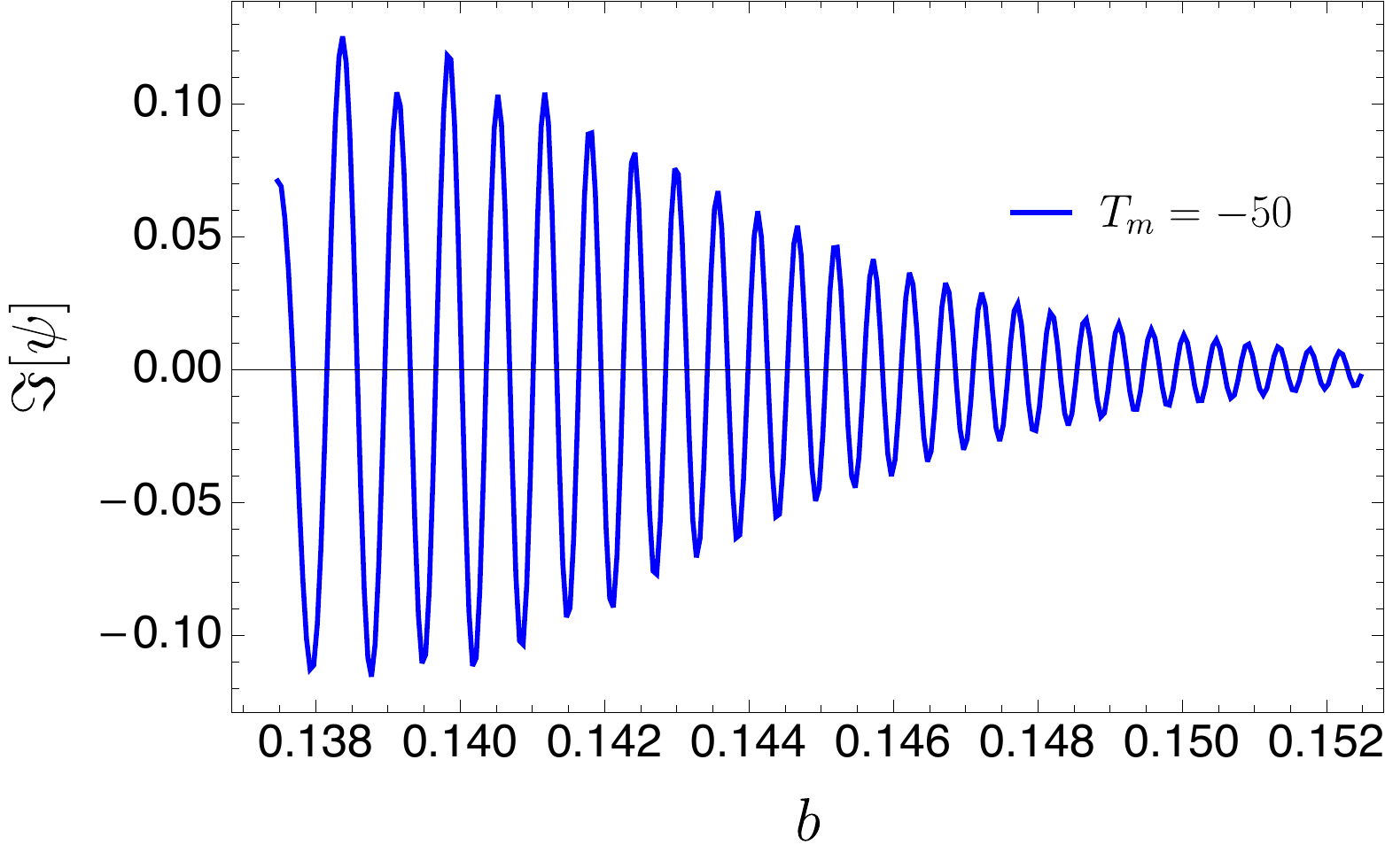}
\includegraphics[scale=0.53]{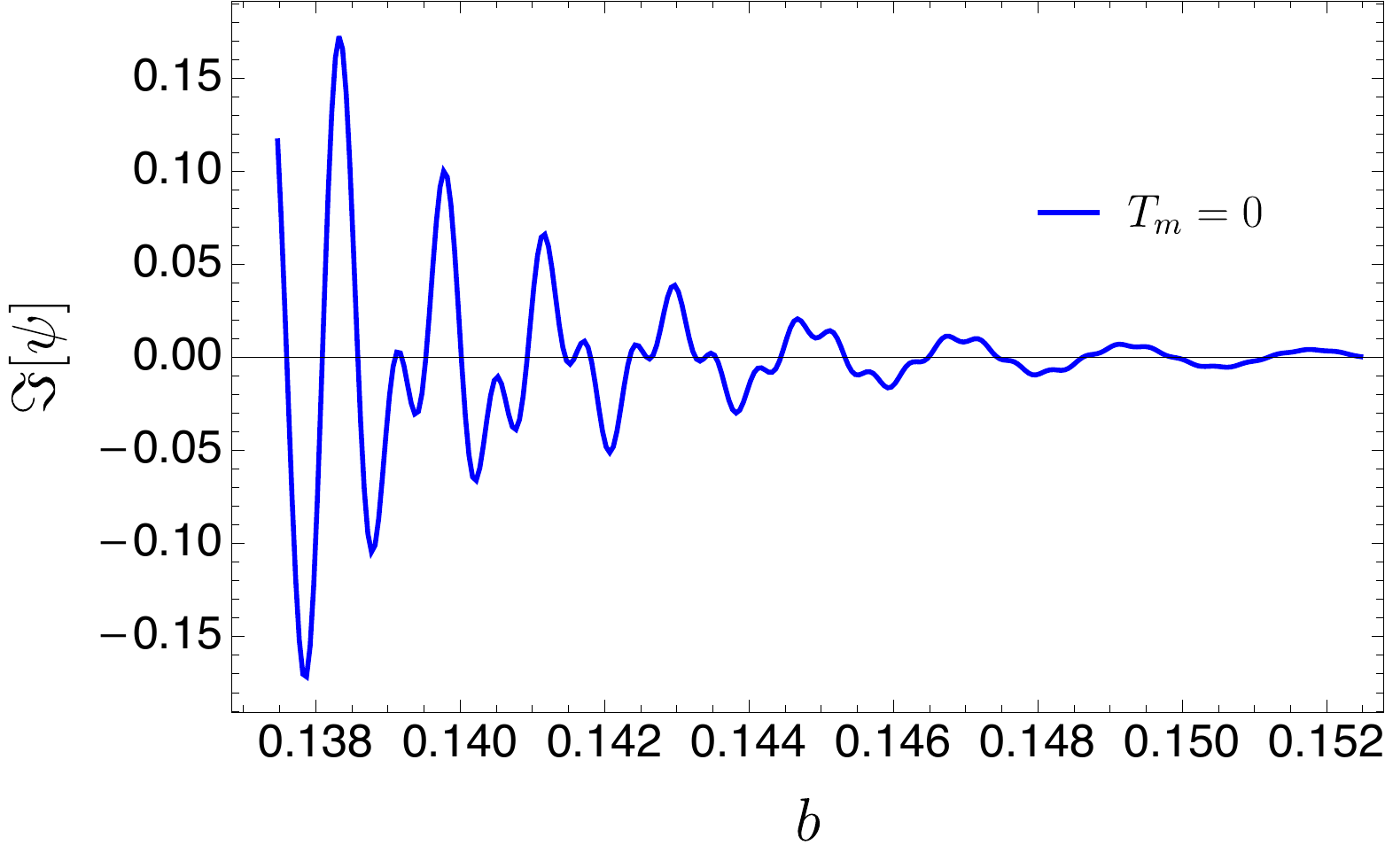}
\includegraphics[scale=0.53]{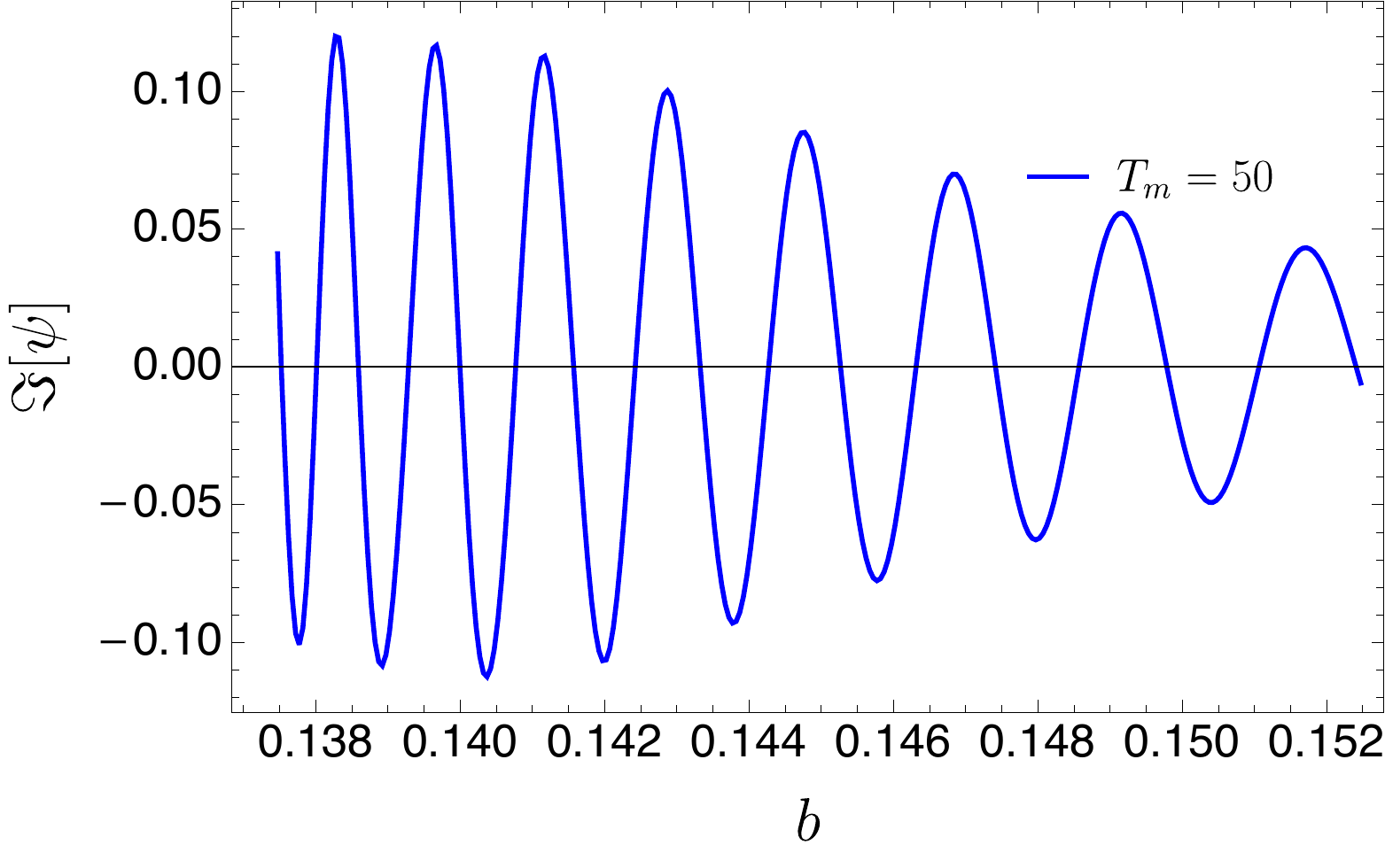}
\caption{Imaginary part of the wave function in the classically allowed region, for $T_m=-50,0,50$ respectively. Here we consider a "matter" wave packet with $\sigma_{Tm}=30$. We also have $k=0$, $m=1$ and $\phi=10^6$. The bounce in $b$ space happens at $T_m=0$.} 
\label{figpsit}
\end{figure}

\begin{figure}[H]
\centering
\includegraphics[scale=0.53]{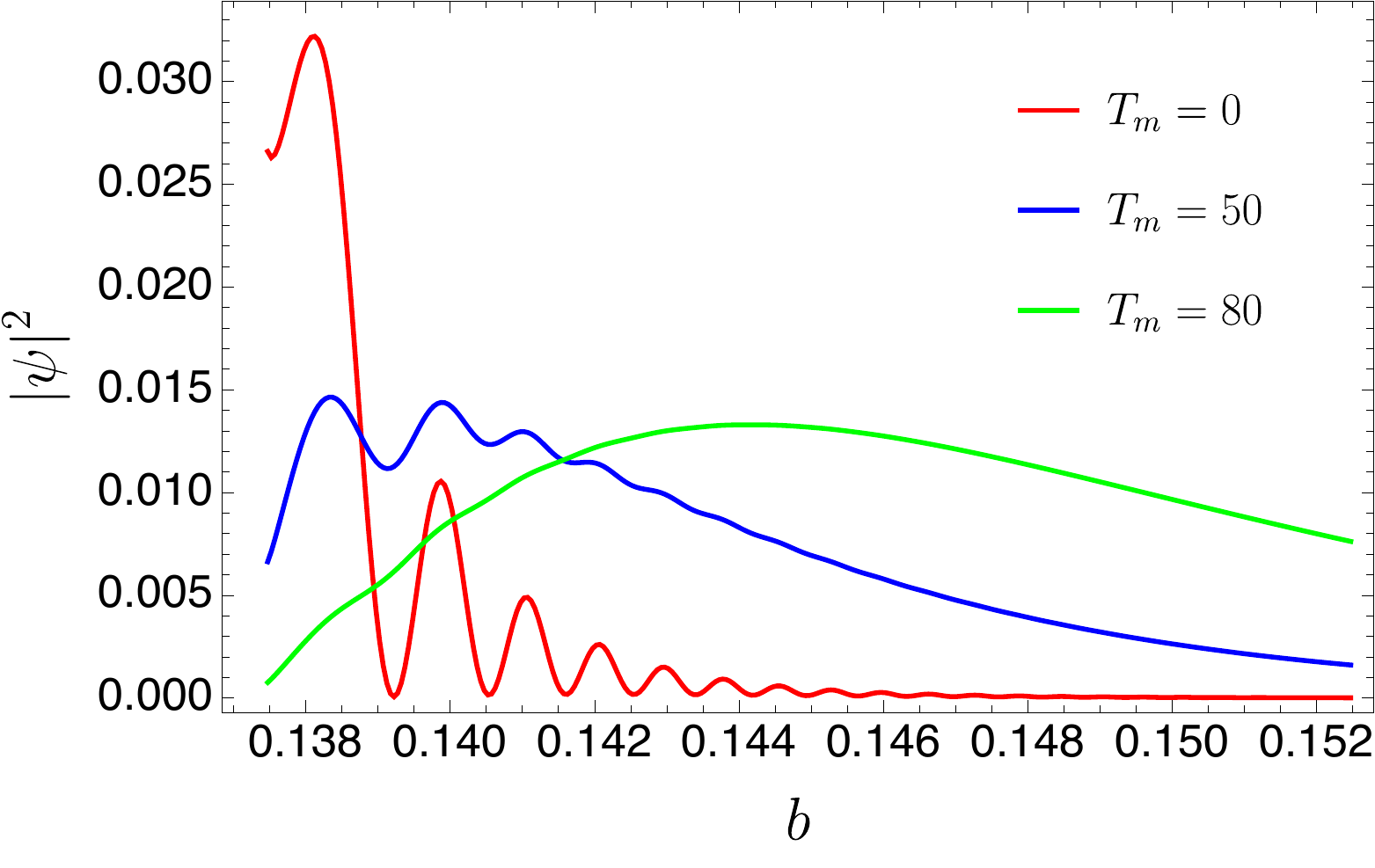}
\caption{The module squared of the wave function in the classically allowed region, for $T_m=0,50,80$ respectively. We consider a "matter" wave packet with $\sigma_{Tm}=30$ and set  $k=0$, $m=1$ (central value) and $\phi=10^6$. The phenomenon of ringing is in evidence. Although the internal beats of the incident and reflected packets are separately invisible in the separate $|\psi|^2$, they become visible near the reflection due to the interference of the incident and reflected wave.  } 
\label{figpsiab}
\end{figure}

We now illustrate the behaviour of the wave packets in the region $b>b_*$, focusing first on the matter clock, $T_m$. The waves, therefore, have the form:
\begin{widetext}
\begin{eqnarray}
\psi_{\pm}(b,T_m)=e^{\frac{i}{\plk}(P_{\pm}(b,\boldsymbol{\alpha}_0)-m^2T_m)}\frac{1}{(2\pi\sigma_{Tm}^2)^{1/4}}\exp\left[-\frac{(X^{\text{eff}}_{\pm m}(b)-T_m)^2}{4\sigma_{Tm}^2}\right],
\label{eqpsitm}
\end{eqnarray}
\end{widetext}
with
\begin{eqnarray}
X^{\text{eff}}_{\pm m}(b)=\int^b_{b_*}\frac{2\phi V}{9m^2x}\sin\left(\frac{2\arctan x}{3}\pm\frac{2\pi}{3}\right)db.
\end{eqnarray}
In Figure \ref{figpsit} we plot the imaginary part\footnote{This is merely for definiteness and plot clarity. The real part has a similar behaviour, out of phase with the imaginary part.}  of the wave packet at three different matter times, $T_m$, before, at and after the bounce. We see that just as in \cite{gielen}, well before and after the bounce,  the wave function behaves as  separate envelopes $\psi_+(b,{T}_m)$ and $\psi_-(b,{T}_m)$, respectively. These packets correspond to  matter and Lambda domination. Near the bounce, however, the two packets interfere, resulting in the ``ringing'' phenomenon. 
This interference is transposed into the amplitude of the total wave function, as observed in figure \ref{figpsiab}, where we can see the small oscillations in the distribution, which fade away far from the bounce time.
\begin{figure}[H]
\centering
\includegraphics[scale=0.53]{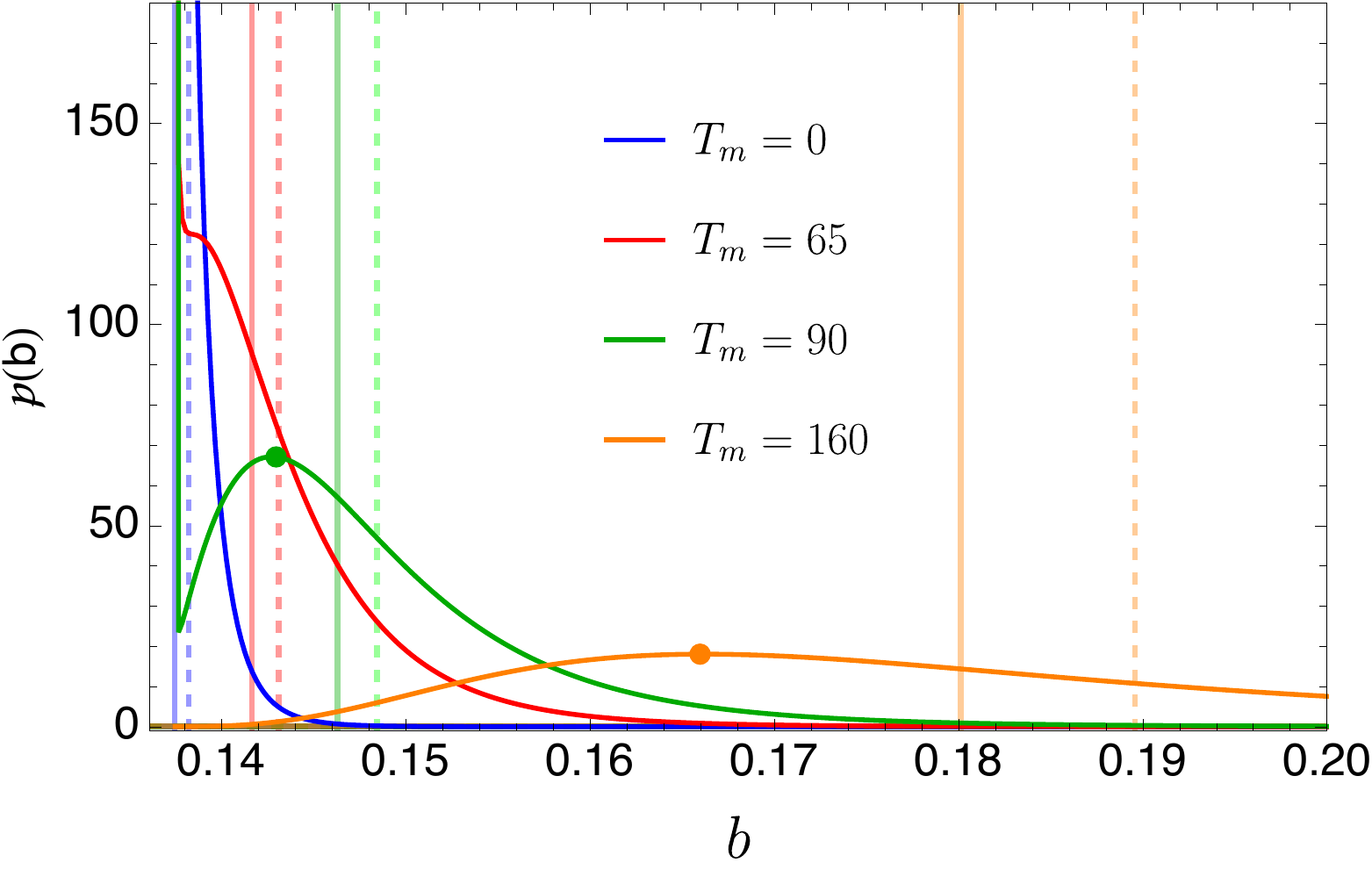}
\caption{Probability calculated using the semiclassical measure for $T_m=0,65,90,190$ respectively, for a ``matter'' $m$ wave packet with associated $\sigma_{Tm}=30$. We set $k=0$, $m=1$ (central value) and $\phi=10^6$. For each time (i.e. for any given  color) the point marks the peak of the distribution
(where present), the dashed vertical line the average $b$ and the solid vertical line the position of $b$ for the classical trajectory for that  time. As we see the peak and the average of the distribution both disagree substantially with the classical trajectory, in opposite directions. }
\label{figprob}
\end{figure}

This illustrates a phenomenon present in general, well beyond the saddle approximation used in obtaining these wave functions. 
However, we may consider using the semiclassical measure \cite{gielen}, inferred from the same approximation used in computing these wave packets. Then, the probability is given by
\begin{eqnarray}\label{semiclassprob}
\mathcal{P}_i(b,T_i)=|\psi_{+i}|^2|\partial_bX^{\text{eff}}_{+i}|+|\psi_{-i}|^2|\partial_bX^{\text{eff}}_{-i}|.
\end{eqnarray}
Applying this to the matter case we have
\begin{eqnarray}
\partial_bX^{\text{eff}}_{\pm m}=\frac{2\phi V}{9m^2x}\sin\left(\frac{2\arctan x}{3}\pm\frac{2\pi}{3}\right)
\end{eqnarray}
and $\psi_{\pm i}$ are given by (\ref{eqpsitm}). This probability is plotted in Fig.~\ref{figprob} for different times. As we can see, we no longer have the ringing effect near $T_m=0$ in this approximation.  Nevertheless, we find the presence of a divergence as $b\rightarrow b_*$, since
\begin{eqnarray}
\mathcal{P}_m(b\rightarrow b_*,T_i)=\frac{2\phi V_*}{9m^2x(2\pi\sigma_{Tm}^2)^{1/2}}e^{-\frac{T_m^2}{2\sigma_{Tm}^2}}.
\end{eqnarray}
This divergence becomes irrelevant (exponentially suppressed)  for $|T_m|\gg\sigma_{Tm}$ due to the exponential factor, as observed in the plot for $T_m=190$. For smaller $T_m$ a double peaked distribution occurs, the main peak corresponding to the Gaussian distribution and the other one associated with the semiclassical measure factor. Approaching the bounce at $T_m=0$, the Gaussian peak vanishes (in figure \ref{figprob} the transition point is around $T_m=65$) and the distributions remain only peaked around $b=b_*$.

We now examine the evolution of the width of the peak with time far away from the bounce ($b\gg b_*$). In this limit, the expressions for $X^{\text{eff}}$ take the form:
\begin{eqnarray}
X_{+m}^{\text{eff}}\approx -\frac{1}{3b^3},
\end{eqnarray}
\begin{eqnarray}
X_{-m}^{\text{eff}}\approx -\frac{\sqrt{27}}{4b_*^3}\ln b.
\end{eqnarray}

The standard deviation in $b$ space can be obtained through $\sigma_X=\sigma_T$ with 
\begin{eqnarray}
\sigma_b=\bigg |\frac{db}{dX^{\text{eff}}}\bigg|\sigma_X,
\end{eqnarray}
which leads to 
\begin{eqnarray}
\frac{\sigma_b}{b}=b^3\sigma_{Tm}
\end{eqnarray}
for $X_{+m}^{\text{eff}}$ and 
\begin{eqnarray}
\frac{\sigma_b}{b}=\frac{4b_*^3}{\sqrt{27}}\sigma_{Tm}
=\frac{2m}{\sqrt{\phi}}\sigma_{Tm}
\end{eqnarray}
for  $X_{-m}^{\text{eff}}$.
Therefore, deep in the matter epoch, i.e. for $T_m\gg 0$, since $b$ increases with $T_m$ the peak of the wave packet becomes larger with $b^3$, which demonstrates the quantum nature of the early universe. On the other hand, for $T_m \ll 0$, we are in the Lambda epoch and $b$ increases with $|T_m|$. However, the standard deviation does not depend on $b$ and so the width of the distribution stays approximately constant over time, corresponding to the propagation of a semi classical state.

\subsection{LAMBDA CLOCK}

\begin{figure}
\centering
\includegraphics[scale=0.53]{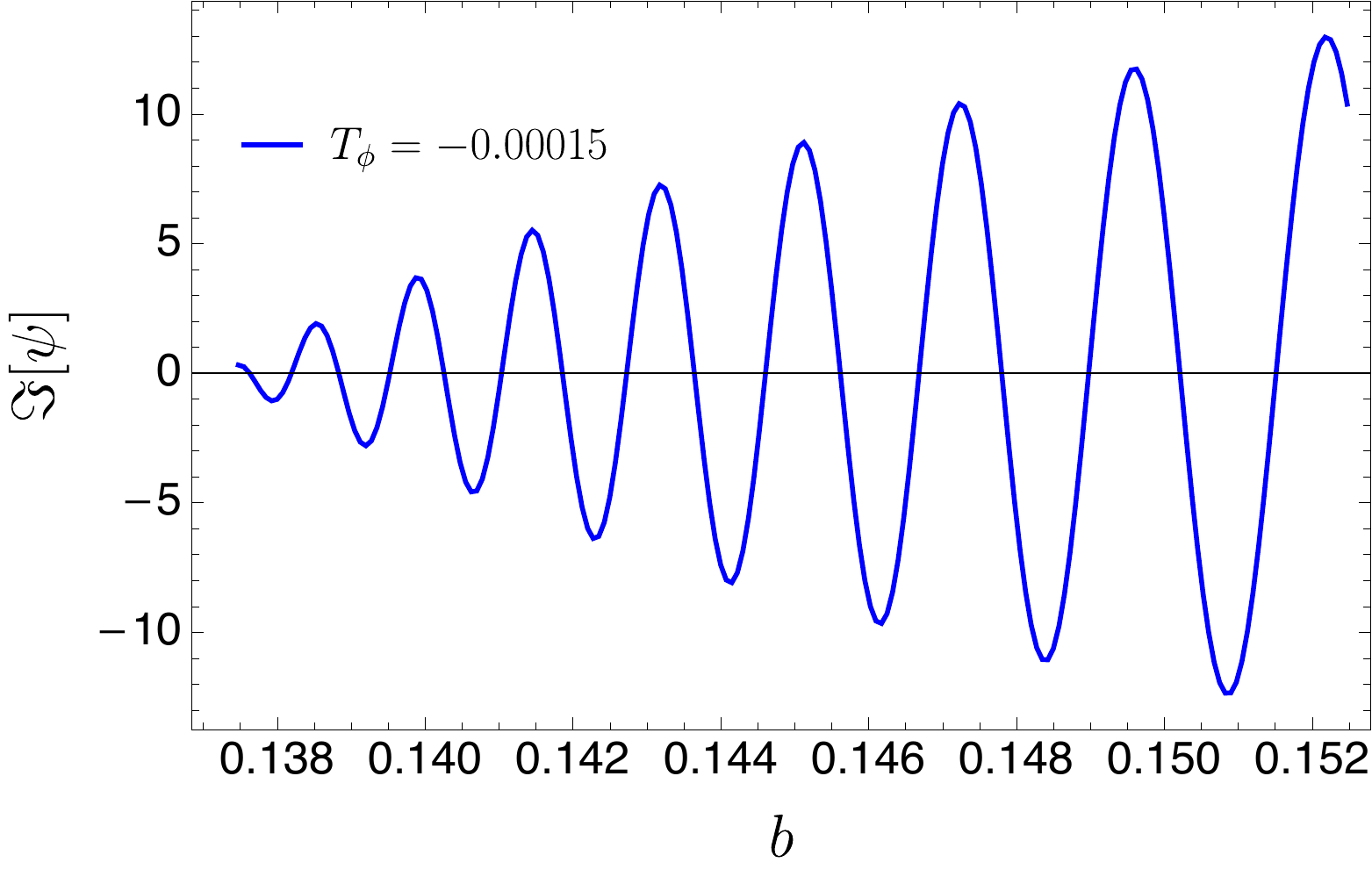}
\includegraphics[scale=0.53]{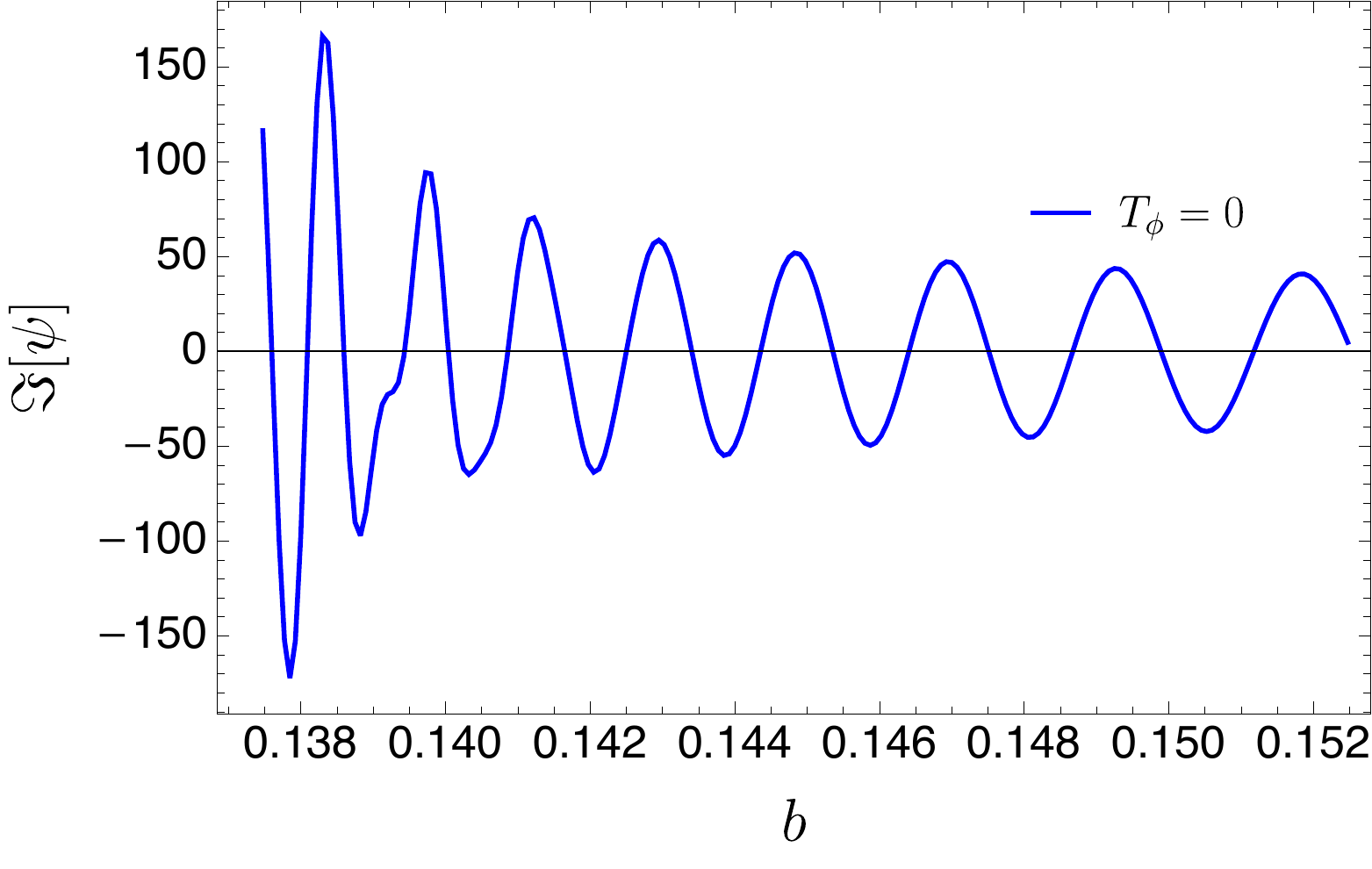}
\includegraphics[scale=0.53]{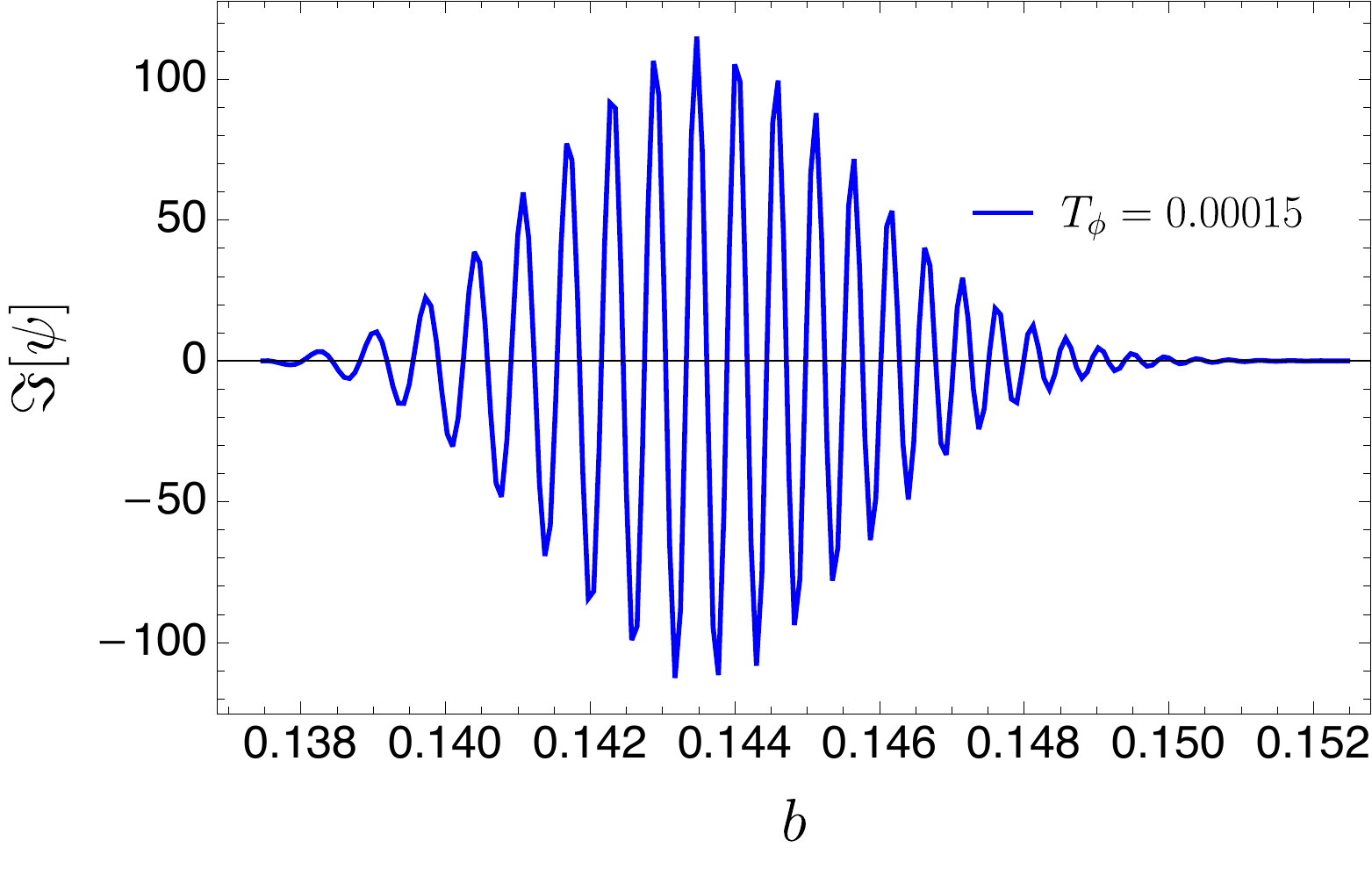}
\caption{Imaginary part of the wave function in the classically allowed region, for $T_{\phi}=-0.00015, 0, 0.00015$ respectively. Here we consider a "Lambda" wave packet with $\sigma_{T\phi}=0.00003$. We also have $k=0$, $m=1$ and $\phi=10^6$. The bounce in $b$ space happens at $T_{\phi}=0$.}
\label{figpsitl}
\end{figure}
\begin{figure}
\centering
\includegraphics[scale=0.53]{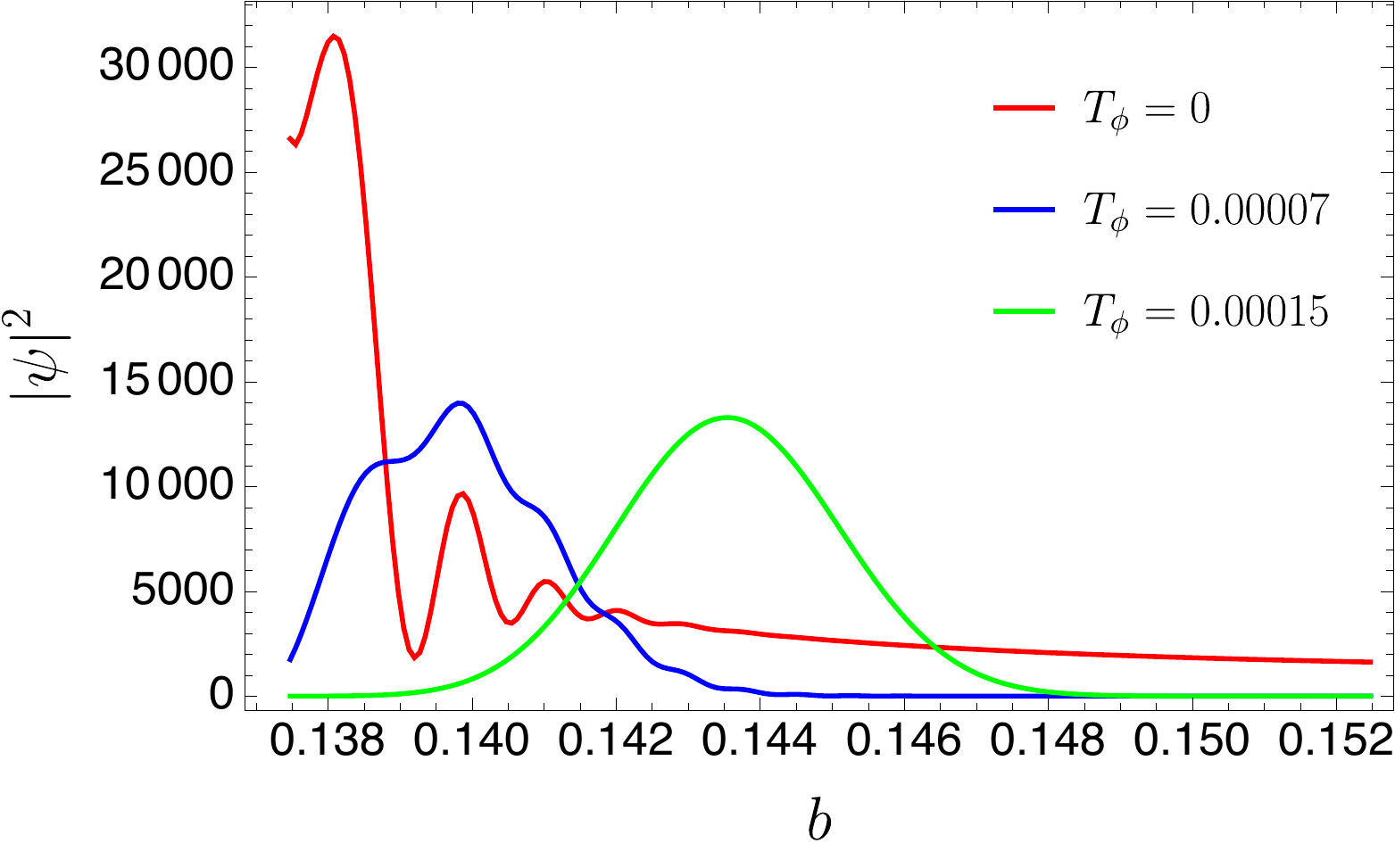}
\caption{Module squared of the wave function in the classically allowed region, for $T_{\phi}=0,0.00007,0.00015$ respectively. Here we consider a "Lambda" wave packet with $\sigma_{T\phi}=0.00003$. We also have $k=0$, $m=1$ and $\phi=10^6$. The bounce in $b$ space happens at $T_{\phi}=0$.}
\label{figpsiabl}
\end{figure}

\begin{figure}
\centering
\includegraphics[scale=0.51]{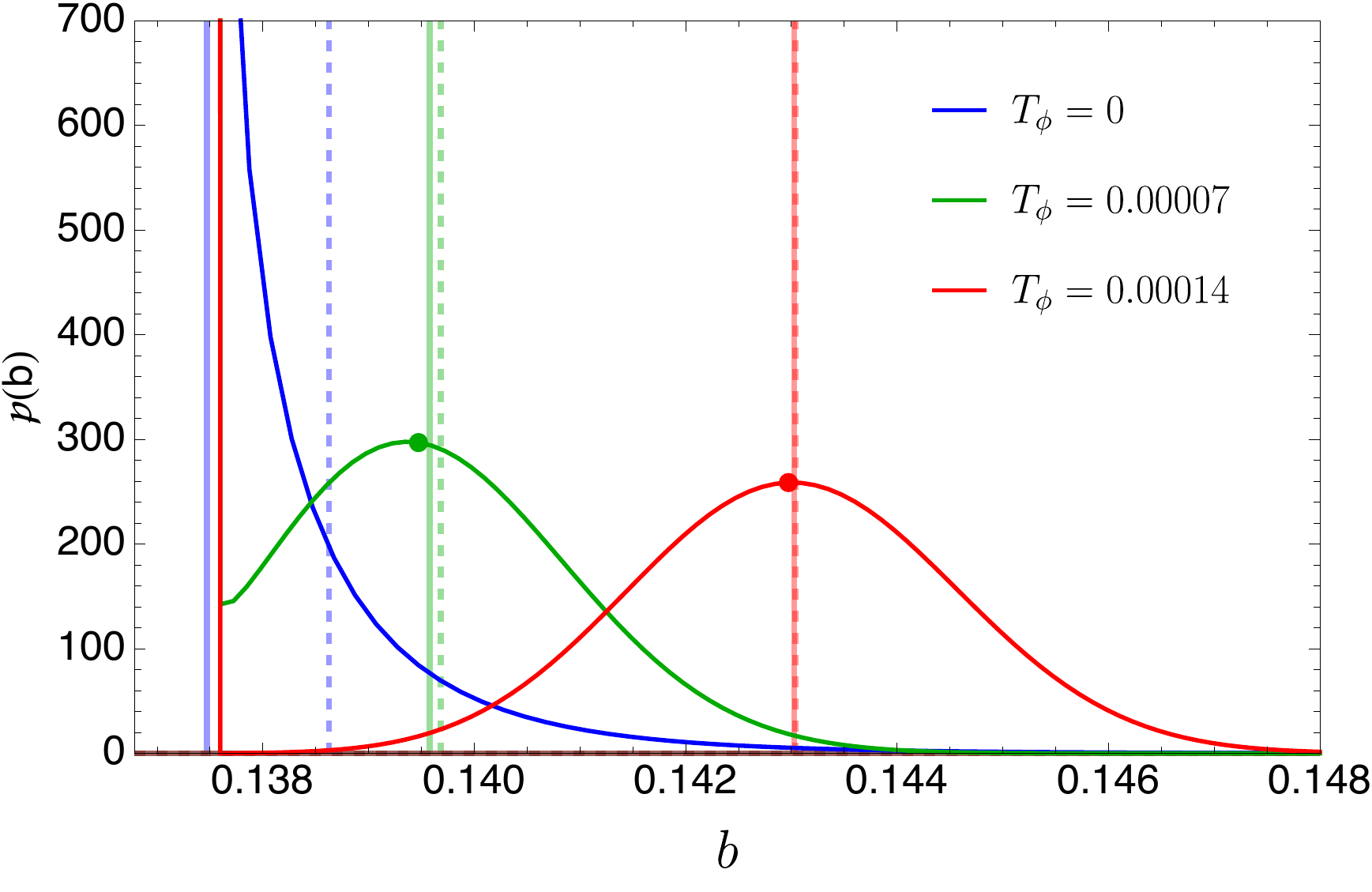}
\caption{
Probability calculated using the semiclassical measure for $T_{\phi}=0,0.00007,0.00014,0.0003$, respectively, for a Lambda wave packet with associated $\sigma_{T\phi}=0.00003$. We set $k=0$, $m=1$ and $\phi=10^6$ (central value). For each time (i.e. for any given  color) the point marks the peak of the distribution
(where present), the dashed vertical line the average $b$ and the solid vertical line the position of $b$ for the classical trajectory for that  time. As we see the peak and the average of the distribution both disagree substantially with the classical trajectory, in opposite directions. Near the bounce the average $b$ is substantially higher than its classical value, as the blue curve illustrates.} 
\label{figpsiablc}
\end{figure}

If we use the Lambda clock the wave function is given by:
\begin{widetext}
\begin{eqnarray}
\psi_{\pm}(b,T_{\phi})=e^{\frac{i}{\plk}(P_{\pm}(b,\boldsymbol{\alpha}_0)-\phi T_{\phi})}\frac{1}{(2\pi\sigma_{T\phi}^2)^{1/4}}\exp\left[-\frac{(X^{\text{eff}}_{\pm \phi}(b)-T_{\phi})^2}{4\sigma_{T\phi}^2}\right],
\label{eqpsitphi}
\end{eqnarray}
\end{widetext}
with
\begin{eqnarray}
X^{\text{eff}}_{\pm \phi}(b)=\frac{P_\pm}{\phi}-\int^b_{b_*}\frac{2 V}{9x}\sin\left(\frac{2\arctan x\pm 2\pi}{3}\right)db.
\end{eqnarray}
We observe a similar behaviour in the imaginary part of the wave function (Figure \ref{figpsitl}) as compared to the previous subsection.
We plot again $|\psi|^2$ at different $T_{\phi}$  in Figure \ref{figpsiabl}, where we can observe the ``ringing'' phenomenon during the bounce. 
This disappears using the semiclassical measure (Figure \ref{figpsiablc}), but corrections are still visible around the bounce.
Specifically, we see that the peak of the distribution tends to be biased to lower values with respect to the classical trajectory. 
This can be derived analytically in regions where the one of the waves dominates (incident or reflected) and is due to the measure effect.
This not only induces a divergence at $b=b_\star$ but also induces a correction in the motion of the peak, as can be obtained by examining the derivative: 
\begin{widetext}
\be
X^{\rm eff }_{\pm\phi}- T_\phi= -\frac{2\sigma_{T\phi}^2b}{3|\partial_bX_{\pm\phi}^{\text{eff}}|^2}\left[4\cos^2\left(\frac{\arctan x\pm\pi}{3}\right)+\left(\frac{b^6}{b_*^6x^3}-\frac{8}{3x}\right)\sin\left(\frac{2\arctan x\pm2\pi}{3}\right)-\frac{2}{3x^2}\cos\left(\frac{2\arctan x\pm2\pi}{3}\right)\right].
\ee
\end{widetext}
In contrast the average value of $b$ tends to be higher than the classical value around the bounce.
This is likely to be the relevant prediction of this calculation. For this reason we will examine this effect in more detail 
in the next Section, with real observational data. 

We close this Section by noting that for $b\gg b_*$ we obtain the Lambda limit from $X_{-\phi}^{\text{eff}}$
\begin{eqnarray}
X_{-\phi}^{\text{eff}}\approx \frac{b^3}{3}
\end{eqnarray}
and the matter limit from $X_{+\phi}^{\text{eff}}$
\begin{eqnarray}
X_{+\phi}^{\text{eff}}\approx C_{\phi}+\frac{32}{6561}\frac{b_*^{12}}{b^9}= 
C_{\phi}+\frac{2}{9}\frac{m^4}{\phi^2 b^9},
\end{eqnarray}
with $C_{\phi}$ a constant. We can now analyse the behaviour of the peak far away from the bounce in each epoch.
For $X_{+\phi}^{\text{eff}}$ we get the standard deviation in $b$ given by
\begin{eqnarray}
\frac{\sigma_b}{b}=\frac{729}{32}\frac{b^9}{b_*^{12}}\sigma_{T\phi}=\frac{\phi^2}{2m^4}\sigma_{T\phi}b^9.
\end{eqnarray}
As we can see, the width of the peak will increase with the power $b^9$ and so, deep in the matter epoch, a long time before the bounce ($T_{\phi}\ll 0$), the wavepacket will become larger since $b$ increases with $|T_{\phi}|$.
The opposite happens in the Lambda epoch since for $X_{-\phi}^{\text{eff}}$ we have:
\begin{eqnarray}
\frac{\sigma_b}{b}=\frac{\sigma_{T\phi}}{b^3}
\end{eqnarray}
and $b$ increases with $T_{\phi}$, therefore the wave packet becomes more narrow in $b$ space as time passes.

\section{Phenomenology around the transition region}\label{phenomenology}
We now make use of the fact that we happen to be living near the transition from deceleration to acceleration, to examine possible observational effects of a bounce in connection space. Indeed, no single component dominates the recent Universe, with the present density parameter associated with Lambda given by \cite{planck}:
\begin{eqnarray}\label{OmegaL0}
\Omega_{\Lambda 0}=
\frac{\Lambda}{3H_0^2}=
\frac{1 }{\phi H_0^2}
=
0.6889\pm 0.0056
\end{eqnarray}
(where the subscript $0$ denotes ``now'', whereas, we recall, * denotes the bounce point). The present Hubble parameter and value of $b$ (with convention $a_0=1$) is:
\begin{equation}\label{H0}
H_0=b_0=\sqrt{m+\frac{1}{\phi}}
=67.37\pm 0.54 {\rm km} \, {\rm s}^{-1}.
\end{equation}
From (\ref{OmegaL0}) and (\ref{H0}) we can solve for $m$ and $\phi$ in terms of $\Omega_{\Lambda 0}$ and $H_0$, and then use  (\ref{eqastar}) and (\ref{eqbstar})
to find:
\begin{eqnarray}
a_*=\frac{(1-\Omega_{\Lambda 0})^{1/3}}{(2\Omega_{\Lambda 0})^{1/3}}.
\end{eqnarray}
\begin{eqnarray}
b_*=\left(\frac{27}{4}\right)^{1/6}(1-\Omega_{\Lambda 0})^{1/3}\Omega_{\Lambda 0}^{1/6}H_0
\end{eqnarray}
or the central values:
\begin{eqnarray}
b_*&=&6.027\times 10^{-11} {\rm y}^{-1}\\  z_*&=&\frac{1}{a_*}-1=0.642.
\end{eqnarray}
Converting into more frequently used variables (using standard numerical codes for relating $a$ and $t$) we find that the bounce occurs when:
\begin{eqnarray}
H_\star&=&{ \sqrt {3\Omega_{\Lambda 0}}} H_0= 96.85 \, 
{\rm km}\;{\rm  s}^{-1} {\rm Mpc}^{-1}\\
t_\star&=& 7.673 \, {\rm Gy} .
\end{eqnarray}
We are thus close to the bounce, both in terms of metric (redshift, $z_\star$) and connection $b$. Specifically, $b_\star/b_0\approx 0.875$ so that:
\begin{equation}
    \frac{b_0-b_\star}{b_\star}\approx 0.142.
\end{equation}
Whether we are close enough to justify a power-law expression for our predictions will presently be assessed.

\subsection{Power series expansions near the bounce}
Near the reflection point, we can approximate:
\bea\label{bexpansion}
X^{\rm eff}_{\pm \phi}&=&
\sum_n (\pm )^n \alpha^X_n b_\star ^{3-n/2} (b-b_\star)^{n/2} 
\eea
and refer to computer algebra to find as many coefficients as needed\footnote{ Concretely, $\alpha_1^X= - \frac{\sqrt{2}}{9}$,
$\alpha_2^X= \frac{11} {27} $,
$\alpha_3^X= - \frac{137}{162\sqrt{2}} $,
$\alpha_4^X= \frac{131} {243} $,
$\alpha_5^X= - \frac{8761}{19440\sqrt{2}} $,
$\alpha_6^X= \frac{329} {2187} $, etc.}.
In general, the $\Lambda$ probability factor is given by:
\bea
{\cal P}_\phi (b;T_\phi)&=&\frac{e^{-\frac{ (X_{+  \phi}^{\rm  eff} -T_\phi)^2}{2 \sigma_{T\phi} ^2}}|\partial_bX^{\text{eff}}_{+\phi}|
+e^{-\frac{ (X_{- \phi}^{\rm  eff} -T_\phi)^2}{2 \sigma_{T\phi} ^2}}
|\partial_bX^{\text{eff}}_{-\phi}|}{\sqrt{2\pi\sigma^2_{T\phi}}}.\nonumber
\eea
Using only the first term in (\ref{bexpansion}) this reduces to:
\be
{\cal P}_\phi (b;T_\phi) \approx
\frac{e^{-\frac{ (X_\star -T_\phi)^2}{2 \sigma_{T\phi} ^2}}
+e^{-\frac{ (X_\star +T_\phi)^2}{2 \sigma_{T\phi} ^2}}
}{\sqrt{2\pi\sigma^2_{T\phi}}}|\partial_bX_\star |
\ee
where $X_\star(b)=\frac{\sqrt{2}}{9}b_\star^{5/2}(b-b_\star)^{1/2}$. 
Hence, in this approximation, we have that the distribution is not a Gaussian, but an exponential distribution (i.e. of the form $\sim \exp(-\lambda (b-b_0)) $), before we consider the divergence due to the measure factor $|\partial_bX_\star |$ (this is corroborated by Fig.~\ref{figpsiablc}). In this approximation the method of the 
images for reflections holds true~\cite{interfreflex}, that is, the total wave is the sum of the incident wave and an identical mirror wave.
However none of these two facts are true if we add more than the first term of (\ref{bexpansion}). This is because the reflected and incident wave propagate in different dispersive media. 
Nonetheless the first term makes the point that the distribution is very skewed and squashed against $b_\star$.

We will use the average $\bar b$ as an indicator of the quantum anomalies that might be observed. This
can be computed directly in terms on $X^{\rm eff}_\phi$ because:
\bea\label{aveb}
\bar b(T_\phi)&=&\int_{b_\star}^\infty db\,  P(b,T_\phi) b\nonumber\\
&=&\int_{-\infty}^\infty dX^{\rm eff}_\phi P(X^{\rm eff}_\phi,T_\phi ) b(X^{\rm eff}_\phi),
\eea
where we drop the $\pm$ subscript in $X^{\rm eff}_\phi$, since this is implicit in the domain.
Then $P(X^{\rm eff}_\phi,T_\phi)$ is a normal distribution in $X^{\rm eff}_\phi$ centered at $X^{\rm eff}_\phi=T_\phi $ with variance $\sigma_{T\phi}$, and if we approximate $b(X^{\rm eff}_\phi)$ via the Taylor expansion
following from (\ref{bexpansion}):
\bea
b&=& b_\star+
\sum_{n=2}^\infty \alpha_n^b \frac{(X^{\rm eff}_\phi)^n}{b_\star^{3n-1}} \nonumber\\
&\approx& b_\star +\frac{81}{2 b_\star^5} (X^{\rm eff}_\phi)^2
-\frac{2673 }{2 b_\star^8} (X^{\rm eff}_\phi)^3
+ \frac{85293 }{2 b_\star^{11}} (X^{\rm eff}_\phi)^4+...\nonumber
\eea
we reduce the problem to the moments of a Gaussian distribution. These are $s_n=0$, for odd $n$, $s_n=(n-1)!!$ for even $n\ge 2$ (plus the normalization condition $s_0=1$).
We thus have:
\begin{eqnarray}\label{pertavb}
\frac{ \bar b}{b_\star} &=& 1 +\sum_{n=2}^\infty \alpha_n^b\frac{\sigma_{T\phi}^n}{b_\star ^{3n}}\sum_{k=0}^{n}\binom{n}{k}s_k\left(\frac{T_\phi}{\sigma_{T\phi}}\right)^{n-k}\nonumber\\
 &=&1 +
 \sum_{n=2}^\infty \sum_{k=0}^{n}
 \alpha_n^b s_k \binom{n}{k}
 \left(\frac{\sigma_{T\phi}}{b_\star^3}\right)^k
 \left(\frac{T_\phi}{b_\star^3}\right)^{n-k}
\end{eqnarray}
resulting in a double expansion: in $T_\phi$ (resulting from our expansion of $b$ around $b_\star$) and in even powers of $\sigma_{T\phi}/b_\star^3$.

This expansion is only suitable when the effect is reasonably small (because $\sigma_{T\phi}/b_\star^3$ is small). For example at $T_\phi=0$ (i.e. at the bounce) we 
have:
\bea\label{corratbounce}
\frac{\bar b-b_\star}{b_\star}&=& 
\sum_{n=1}^\infty \alpha_{2n}^b \frac{\sigma_{T\phi}^{2n}}{b_\star^{6n}} (2n-1)!!.
\eea
It can be numerically checked that this does not converge if $\sigma_{T\phi}/b_\star^3$ is bigger than 
about $0.008$, for which the fractional correction (\ref{corratbounce}) 
is only about $0.3\%$. However, 
when the series converges, its first term is a already a very good approximation. 

Bearing this in mind, we can derive a perturbative expression for the redshift profile of the correction. 
Such an expression is only practical if we are reasonably close to $z=z_\star$. 
Setting  $\sigma_{T\phi}=0$ in (\ref{pertavb}) we obtain a perturbative expansion for the classical trajectory:
\bea\label{bclspert}
\bar b (\sigma_{T\phi}=0)&=& b_{cl}=b_\star+
\sum_{n=2}^\infty \alpha_n^b \frac{T_\phi^n}{b_\star^{3n-1}}
\eea
as expressed by $X^{\rm eff}_\phi=T_\phi$ (cf. (\ref{bexpansion})). 
Keeping  only first order terms in $\sigma_{T\phi}^2/b_\star^6$ 
we have:
\begin{eqnarray}
\bar b&=&b_{cl} +\frac{\sigma_{T\phi}^2}{b_\star ^5}
\sum_{n=2}^\infty \alpha_n^b\frac{n(n-1)}{2} \left(\frac{T_\phi}{b_\star^3}\right)^{n-2}\\
&=&
b_{cl} +\frac{\sigma_{T\phi}^2}{b_\star ^5}
\left(\alpha^b_2+3\alpha^b_3\frac{T_\phi}{b_\star^3}
+ 6 \alpha^b_4 \frac{T_\phi^2}{b_\star^6} +...\right). 
\end{eqnarray}
Using $T_\phi=X^{\rm eff}_\phi(b_{cl})$ and (\ref{bexpansion}) we finally obtain an expansion with leading order terms:
 \begin{equation}
    \frac{\bar b-b_{cl}}{b_{cl}}\approx 
 \frac{b_\star}{b_{cl}}   
    \frac{\sigma_{T\phi}^2}{b_\star^6}\left(
    \frac{81}{2}\mp\frac{891}{\sqrt{2}}\sqrt{\frac{b_{cl}-b_\star}{b_{\star}}}+\frac{9369}{2}
    \frac{b_{cl}-b_\star}{b_{\star}}
   \right) \nonumber
 \end{equation}
 where $\mp$ refers to after/before the bounce (a $-$/$+$).

\begin{figure}
\centering
\includegraphics[scale=0.9]{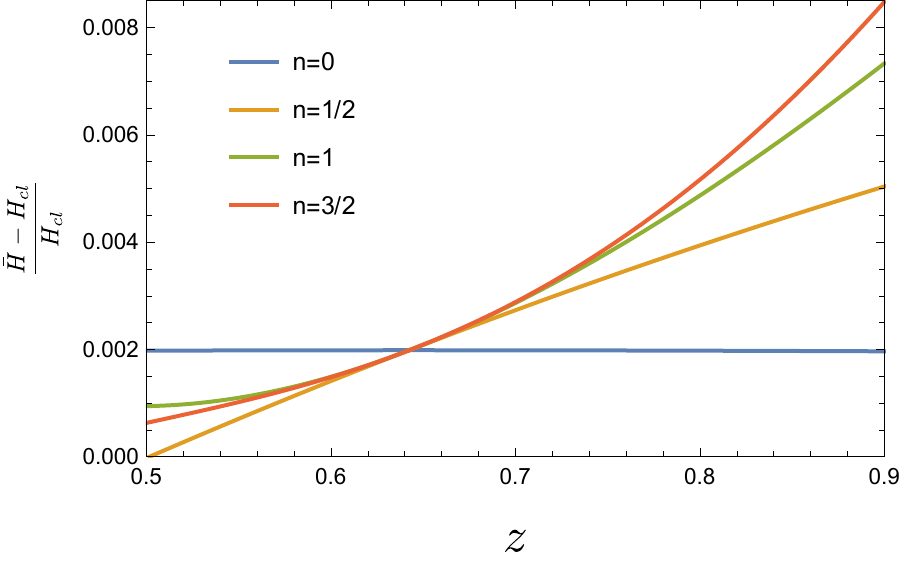}
\caption{Illustration of the correction to the Hubble parameter around $z=z_\star$ 
truncating the series at various orders in $(b_{cl}- b_\star)/b_\star$, for $\sigma_{T\phi}/b_\star^3=0.007$
}
\label{delHn4}
\end{figure}

In Fig.~\ref{delHn4} we see the redshift profile that emerges as we add more and more terms to this expansion. We have evaluated the Hubble parameter assuming that all its uncertainties arise from $b$ and not from the complementary $a$ (as suggested by the analysis in~\cite{BarrowMagueijo,randono}). 
Clearly Taylor expansions are not suitable for the redshift range we need (down to $z=0$), and for values of $\sigma_{T\phi}/b_\star^3$
capable of generating corrections of a few percent (if a connection with the Hubble tension is sought). Nonetheless they could be useful in future settings, and provide good guidance and checks for the numerical results that follow.

%

\subsection{Numerical results}
\begin{figure}
\centering
\includegraphics[scale=0.43]{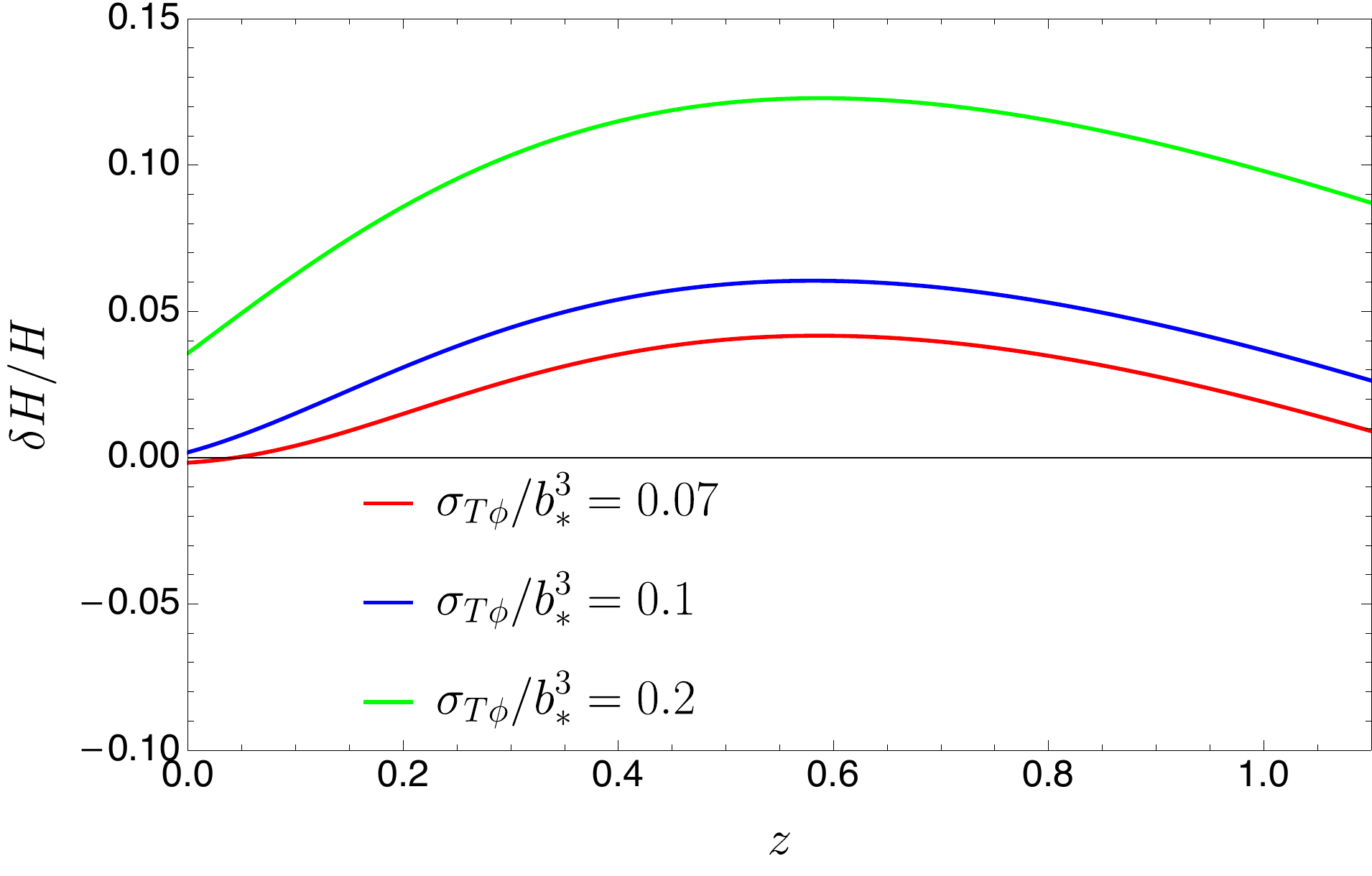}
\includegraphics[scale=0.43]{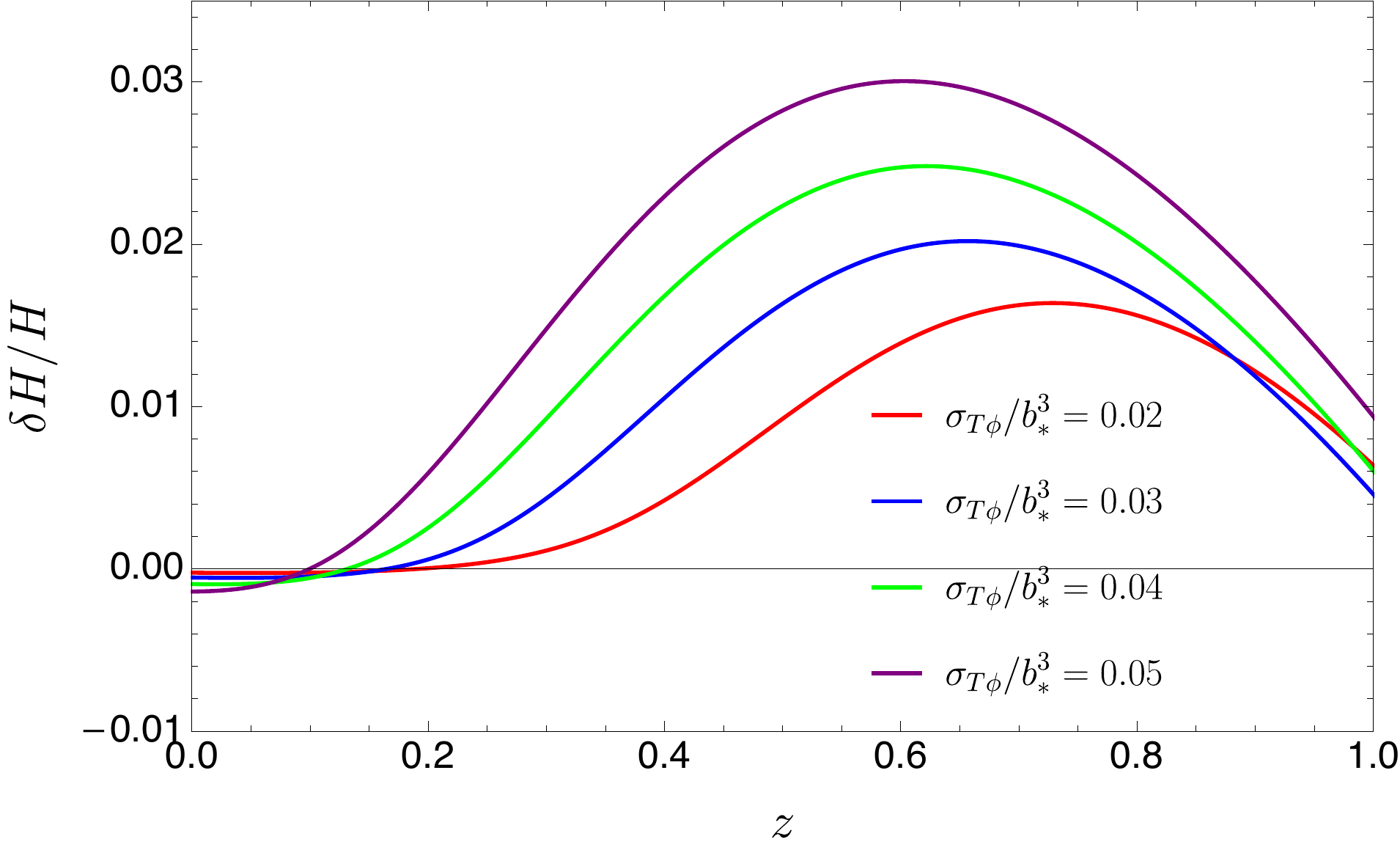}

\caption{Redshift profile of the corrections to the Hubble parameter obtained for various values of $\sigma_{T\phi}/b_\star^3$
large enough to require numerical treatment. As we see the correction can be large enough to explain Hubble tension (top panel), but also more modest (bottom panel). The redshift profile provides testability to this explanation. Note, in particular, how the positive correction around $z\sim z_\star$ may become negative near $z=0$ in some cases. 
}
\label{delHfinal}
\end{figure}

We can also numerically evaluate (\ref{aveb}), check that the results from the code match our analytical results when
$\sigma_{T\phi}/b_\star^3$ is sufficiently small and $z\sim z_\star$, and then push the code beyond the perturbative regime. The outcome of this exercise is plotted in Fig.~\ref{delHfinal}, which is the central result of our paper. As we see, it is not difficult for the average Hubble parameter to be larger than its classical trajectory by 5-10\%. This requires values of $\sigma_{T\phi}/b_\star^3$ of the order of  0.1. Should this be one order of magnitude smaller, the concomitant corrections on $H$ are also roughly one order of magnitude smaller.

It is interesting that the overall effect always implies a {\it larger} measured Hubble parameter for small redshifts, as in the Hubble tension. This is due to the squashing of the wave function due to the reflection, which creates a positively skewed distribution. But by adjusting $\sigma_{T\phi}$, one can obtain a correction as high or as low as required. 

What is truly predictive in our calculation, however, is that the effect has a distinctive redshift profile, as plotted in Fig.~\ref{delHfinal}. The strength of the correction always peaks around $z_\star$, as expected, and as Lambda dominates it subsides.  But as it does so, the correction {\it changes sign}, i.e the average Hubble parameter becomes slightly smaller than its classical prediction. This may happen after $z=0$ (i.e. after nowadays), but as the top panel of Fig.~\ref{delHfinal} shows, 
it is only for the largest values of $\sigma_{T\phi}/b_\star^3$ (and so of the effect near $z=z_\star$) that this happens.

This feature can be checked analytically
 by writing $X^{\rm eff}_{- \phi} \approx X_{CS}=b^3/3$, expanding the resulting $b(X)$ around $X=T_\phi$ (for each $T_\phi$) to second order, and inserting in (\ref{aveb}). This leads to:
\bea
\bar b (T_\phi) &\approx&b_{cl} (T_\phi) +\frac{1}{2}
\frac{\partial ^2 b}{\partial X^2}(T_\phi)\sigma_{T\phi}^2 \\
&= & b_{cl} (T_\phi) -\frac{1}{3^{5/3}}
\frac{\sigma_{T\phi}^2}{T_\phi^{5/3}} \\
&=&b_{cl}(T_\phi)\left(1-\frac{\sigma_{T\phi}^2}{b_{cl}^6}\right).
\eea
The origin of this effect can also be understood qualitatively. It results from the measure factor in the probability in terms of $b$, as given by Eq.~(\ref{semiclassprob}). It shifts the peak and average of the distribution to smaller values than the peak of the Gaussian itself. 

In closing this Section we note that for large values of $\sigma_{T\phi}$, the Lambda wave function will quickly run into the type of behaviour studied in~\cite{gielensing}. This would most likely not be observable, since one expects the matter or radiation clocks to take over in this regime. Should the Lambda clock still be in action deep in the matter epoch, however, this remains an interesting possibility.


\section{Conclusions}
We conclude with an appraisal of the ratio between predictions and free parameters in our model. The free parameters are all in the choice of the quantum state of the Universe. Restricting ${\cal A}({\boldsymbol \alpha})$ to factorizable Gaussian states, there are only two free parameters: the $\sigma_{i}$ (or equivalently, the complementary $\sigma_{Ti}$). The central values $\alpha_{0i}$ are fixed by observations. The larger the $\sigma_{Ti}$ the stronger and more persistent will be the quantum corrections, but obviously these must also comply with constraints. 

It is important to stress that the full class of theories proposed in~\cite{GielenMenendez,JoaoLetter,JoaoPaper}, albeit all equivalent classically, lead to very different quantum theories, subject to different observational constraints, and making different predictions. If there is only one clock (say the Lambda clock\footnote{A set up studied in~\cite{nialambda}, with direct implications for the cosmological constant problem.}) then clearly this paper would not be possible, since it would imply a quantum Universe at redshifts of order one. But if there are various clocks at play, and each one is associated with the different components in the Universe that dominate in different epochs~\cite{JoaoLetter}, then such constraints vanish, since a radiation or matter clock could take over in the matter or radiation epochs\footnote{If there are several clocks at play during the same epoch the situation is once again different~\cite{twoclocks}.}. The best chance for observing quantum effects (other than at the singularity~\cite{gielensing}) would then be at transition regions, where one clock is handing over to another. 

That is precisely the situation studied here, where we have a matter and a Lambda clock. The fact that we have a reflection in connection space at the redshift where Lambda takes over only exacerbates the quantum aspects of the problem. In Section~\ref{wavepackets} we isolated 3 effects that potentially could affect cosmological  observations near the transition from matter to Lambda domination:
\begin{itemize}
    \item Ringing due to the  interference of the incident and reflected waves.
    \item Effects of the semi-classical measure on the trajectory of the peak of the distribution. 
    \item Bias towards higher averages due to the ``squashing against the wall'' effect of the distribution at the bounce. 
\end{itemize}
The first effect would require the wave function to be sufficiently non-Gaussian for the semi-classical measure to break down.
The second effect is more conservative in its assumptions on the wave function, but is likely to be very small. The third is the predominant correction, and we have focused on it in Section~\ref{phenomenology}. 

In this paper we raised the possibility that (at least part of) the Hubble tension might be due to a quantum effect (see~\cite{HubbleTension} for an extensive review, and~\cite{savas},
for example, for the problems in attending to all issues). 
We found that because of the squashing of the wave function during the reflection, the overall probability distribution is very skewed, pushing the average value of the Hubble parameter above the classical prediction for redshifts of order $z=z_\star\approx 0.64$, where the bounce occured. For generic parameters, the effect likes to be small (of the order of 0.1\%) but it is not difficult to push  $\sigma_{T\phi}$ to values where 5-10\% corrections are possible. But what makes our result truly predictive is that we obtain a distinctive profile for the corrections of the Hubble parameter with redshift: higher than the classical prediction around $z\sim z_\star$, but then slightly lower than its classical value when $z\approx 0$. This is depicted in Fig.~\ref{delHfinal}, and it is the central result of this paper. 

It might seem strange that quantum cosmology, often confined to the Planck epoch~\cite{HH,vil-PRD,vil-rev,CSHHV}, appears to be relevant at late times. In part this is due to the pragmatical resolution~\cite{GielenMenendez,JoaoLetter,JoaoPaper} to the problem of time (e.g.~\cite{IshamRovelli}) adopted here. The idea that the late-time Universe might be in the realm of quantum cosmology has been discussed for quite a while. For example~\cite{Kiefer} argued  that, beyond the semi-classical approximation, quantum effects are unavoidable at the turning point of a recollapsing Universe. More recently the matter has received some attention, both in the context of toy models (e.g., Ref.~\cite{QCnow}), or in the context of realistic dark-energy models (see~\cite{vasilev} for a comprehensive review). The latter involve situations where singularities would appear in the classical theory, and revolve around the issue of their resolution or otherwise due to quantum effects (as well as the issue of boundary conditions, in work that resonates with~\cite{GielenMenendez,gielen}). To the best of our knowledge, we are the first to address these issues using the connection representation and in relation to the Hubble tension.

Important questions of interpretation naturally emerge~\cite{Jonathan,Jonathan1}. We would expect the experts in such issues to voice their views. On a more pedestrian level it would be interesting to test our prediction, both directly, and indirectly, for example via the implications to the growth factor of cosmic structure or to the  Integrated Sachs-Wolf effect~\cite{savas}.

\section{Acknowledgments}
We thank Stephon Alexander, Stephen Gielen, Savvas Koushiappas and David Mota for discussions related to this paper. This work was supported by FCT Grant No. 2021.05694.BD (B.A.)  and by the STFC Consolidated Grant ST/T000791/1 (J.M.).



\end{document}